\documentclass[aps,amsmath,amssymb,twocolumn,nofootinbib,superscriptaddress]{revtex4}
\usepackage{graphicx,color}
 \graphicspath{{./}{./figures/}}
\usepackage{hyperref}
\usepackage{dcolumn}
\usepackage{bm}

\definecolor{labelkey}{cmyk}{.4,.2,0,0}

\newcommand{\be}{\begin{equation}}
\newcommand{\ee}{\end{equation}}
\newcommand{\bea}{\begin{eqnarray}}
\newcommand{\eea}{\end{eqnarray}}

\newcommand{\nn}{\nonumber }

\newcommand{\JN}{{\mathbb N}}

\newcommand{\JZ}{{\mathbb Z}}

\newcommand{\JR}{{\mathbb R}}

\newcommand{\ssp}{\hspace{3pt}}
\newcommand{\simlaw}{{\sim_{{\rm in \ law}}}}
\newcommand{\sd}{{\sf d}}

\begin{document}

\title{Diffusion in time-dependent random media and the Kardar-Parisi-Zhang equation}
\author{Pierre Le Doussal} \affiliation{CNRS-Laboratoire
de Physique Th{\'e}orique de l'Ecole Normale Sup{\'e}rieure, 24 rue
Lhomond,75231 Cedex 05, Paris, France}
\author{Thimoth\'ee Thiery}
\affiliation{Instituut voor Theoretische Fysica, KU Leuven}
\date{\today}

\begin{abstract}
Although time-dependent random media with short range correlations 
lead to (possibly biased) normal tracer diffusion, anomalous fluctuations occur away from the most probable
direction. This was pointed out recently in 1D lattice random walks, where 
statistics related to the 1D Kardar-Parisi-Zhang (KPZ) 
universality class, i.e. the GUE Tracy Widom distribution, 
were shown to arise. Here we provide a simple picture for this correspondence, directly in the continuum
as well as for lattice models, which allows to study arbitrary space dimension and to
predict a variety of universal distributions. In $d=1$ we predict and verify numerically
the emergence of the GOE Tracy-Widom distribution for the fluctuations of the transition probability.
In $d=3$ we predict a phase transition 
from Gaussian fluctuations to 3D-KPZ type fluctuations as the bias is increased. We predict
KPZ universal distributions for the arrival time of a first particle from a cloud diffusing in such media.
\end{abstract}
\maketitle


\begin{figure}
\centerline{\includegraphics[width=9cm]{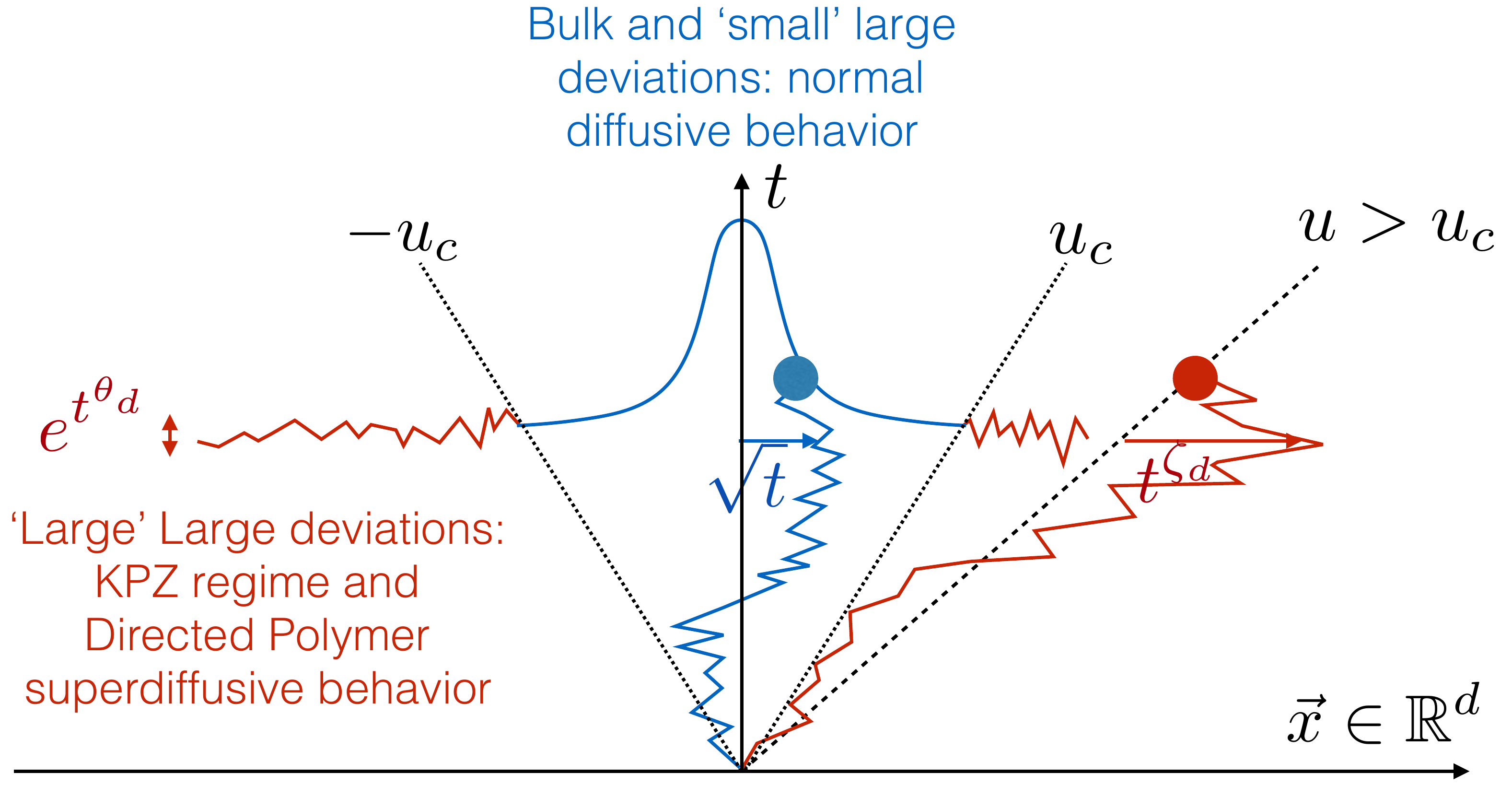}} 
\caption{Conjectural picture for TD-RWRE in arbitrary dimension $d$. While particles tyically diffuse normally as if the random environment had been averaged out, particles conditioned on arriving away from the Gaussian bulk of the distribution in the `large large deviations regime' (for $|x(t)|>u_ct$) are superdiffusive with the roughness exponent of the directed polymer in the pinned phase $\zeta_d > 1/2$. In this regime, fluctuations of the logarithm of the transition probability are large (scale with $t^{\theta_d}$ with $\theta_d = -1 +2 \zeta_d >0$) and identical to those of the height in the rough phase of the KPZ equation. The two phases are separated by an Edward-Wilkinson regime of fluctuations when $x= u_c t + o(t)$. In $d=1,2$ $u_c=0$ while for $d \geq 3$ there is a phase transition with $u_c \neq 0$. The picture is drawn here in the absence of a systematic bias $\vec{f}$. In the presence of a bias the bulk is around $\vec x \sim \vec ft$ and the transition occurs for $\vec x = \vec u t$ with $|(\vec f + \vec u)|=u_c$.}
\label{fig:NicePic}
\end{figure}

{\bf Introduction} Diffusion in random media arises in numerous fields, 
e.g. oil exploration in porous rocks \cite{porous}, spreading of pollutants in inhomogeneous flows
\cite{ParticlesFlows},
diffusion of charge carriers in conductors \cite{bernasconi}, relaxation properties of glasses \cite{trap}, defect motions in
solids, econophysics, population dynamics \cite{BMG2000,Wealth}. 
Many works have studied {\it time independent}, i.e. static, random environments \cite{Review1}, 
in $d=1$ \cite{Static1D} or in higher dimensions, 
with short-range (SR) \cite{StaticD-SR,Lawler} of long-range (LR)
spatial correlations \cite{StaticD-LR}. It was found that static disorder with SR correlations 
is generically irrrelevant above the upper-critical dimension $d_c=2$, leading to normal diffusion in $d=3$,
while LR disorder can lead to anomalous diffusion in any $d$. 

Another important class of random media are time-dependent. These have been studied 
e.g. in the context of wave propagation \cite{BouchaudTimeDep}, dispersion of particles in turbulent flows 
\cite{ParticlesFlows} (the famous Richardson's law \cite{Richardson}), and in the 
problem of the passive scalar \cite{PassiveScalar}. The latter cases involve long range
correlations in the flow, and lead to anomalous transport or multiscaling. The, a priori
more benign, case of SR space-time correlations has received much attention
recently in probability theory, within random walks in time-dependent random environments (TD-RWRE). 
Although then $d_c=0$, and the diffusion is proved to be normal (in a given sample \cite{math1}),
interesting effects were demonstrated, such as a tendency for walkers in the same sample to
coalesce \cite{Nwalkers}, anomalous fluctuations \cite{RAP} and large deviations \cite{LargeDev}. 
Note that TD-RWRE can be generated in a purely static environment by considering
directed 
random walks.

An a priori unrelated topic is stochastic growth 
and the celebrated
Kardar-Parisi-Zhang (KPZ) equation \cite{KPZ}
\be
\partial_t h = \nu_0 \nabla_x^2 h + \frac{\lambda_0}{2} (\vec \nabla_x h)^2 + \sqrt{D_0} \eta \label{KPZ1} 
\ee
where $h(x,t)\in \JR$ is the interface height at time $t$ and point $x \in \mathbb{R}^d$, $\nu$ is the diffusivity, 
$\eta(x,t)$ is the driving noise which, for most of our applications, will be SR space-time correlated. The non-linearity leads to a a non-trivial fixed point and exponents for the scaling of the
fluctuations at large time, i.e. $h(x,t) = v_\infty t + \delta h(x,t)$, with $\delta h \sim t^{\theta_d}
\sim x^{\theta_d/\zeta_d}$ and $\theta_d=2 \zeta_d-1$ from Galilean symmetry
\cite{kardar1987scaling}.
The continuum KPZ equation \eqref{KPZ1} is at the center of a vast universality
class including discrete growth models \cite{PrahoferSpohn2000}, particle transport models
\cite{KrugReview}, dimer covering, directed polymers 
\cite{kardar1987scaling,Johansson2000} and more,
subject in $d=1$ of much recent progress, due to discovery of integrable and determinantal
properties \cite{CorwinReviews}. Beyond exponents
$\zeta_{d=1}=2/3$, the statistics of $\delta h(x,t)$ was shown to be related to the universal Tracy-Widom (TW)
distributions of random matrix theory \cite{TW1994}, with e.g. the GUE (resp. GOE) TW distribution for growth starting from a droplet \cite{droplet} (resp. a flat interface). For general $d$ little is known exactly,
but scaling exponents and universal distributions were obtained numerically in
$d=1,2,3$ \cite{halpin2012-2D,halpin2012-1D,Parisi2016} and, in some cases, 
compared with experiments \cite{halpin-kaz-review}.

Recently, Barraquand and Corwin obtained an exact solution of a discrete TD-RWRE on $\JZ$ with SR correlated jump probabilities, the so-called Beta polymer. The sample to sample fluctuations of the 
logarithm of the cumulative \cite{BarraquandCorwinBeta} and transition \cite{usBeta} probability distribution function (PDF) {\it in the large deviations regime} of the RW, ie. looking away from the most probable direction, were found to be 
distributed with the characteristic KPZ exponent and GUE TW distribution\footnote{Note also the upcoming work \cite{SeppalainenInprep} on the roughness of random walk paths in the Beta polymer model.}. 
This was followed by a proof of the universality of the 1D KPZ equation for the diffusive scaling limit of TD-RWRE on $\JZ$ with weak disorder \cite{CorwinRWREWeakU}. 

These recent results hint at a general connection between TD-RWRE and
KPZ growth. The aim of this Letter is to unveil a simple and general mechanism that explains the appearance of KPZ-type fluctuations in the TD-RWRE problem, 
beyond exactly solvable models, and for general $d$. 
Our main result is that we conjecture the emergence of KPZ fluctuations everywhere in the large deviations regime of TD-RWRE in dimension $d=1,2$, and a {\it a phase transition} in $d\geq 3$ between a low-fluctuations phase for small large deviations and a phase with KPZ class high-fluctuations for large large deviations (see Fig.~\ref{fig:NicePic}). 
We first consider the problem in the continuum setting, and then on the lattice $\JZ^d$. Using the emerging picture, we identify in $d=1$ a natural setting where GOE TW type distribution for the fluctuations of the logarithm of the PDF are expected. This is explicitly checked using simulations of a discrete model.
We finally discuss the emergence of KPZ-related universality in the extreme value statistics of $N \gg 1$ random walker diffusing in the same time-dependent random environment: universality of the PDF of the largest distance travelled by a particle in a cloud of pollutant 
diffusing in a non-homogeneous atmosphere and of the PDF of the first arrival time of the cloud in a given domain.

\medskip

{\bf Main analysis} We consider the Langevin equation for the diffusion of a particle $\vec x(t) \in \JR^d$ in a $d-$dimensional time-dependent random force field $\vec \xi(\vec x,t) + \vec f$, with $\overline{\vec \xi(\vec x,t)}=0$ and $\vec f$ the uniform applied force,
\bea \label{langevin} 
\frac{d}{d t} \vec x(t) = \vec \xi(\vec x(t),t) + \vec f + \vec \eta(t) \ ,
\eea 
with $\vec \eta \in \JR^d$ a thermal Gaussian white noise, $\langle \eta_i(t)   \eta_j(t') \rangle= 2 D \delta_{ij} \delta(t-t')$, and $D$ is the bare diffusion coefficient. Here and below $\langle . \rangle$ refers to the average over thermal fluctuations $\vec \eta$,
and $\overline{(.)}$ over the disorder $\vec \xi(\vec x,t)$.

In a given random environment (i.e. sample) $\vec \xi(\vec x ,t)$ one defines the transition probability 
$P(\vec x_2,t_2|\vec x_1,t_1)$ for a particle which starts at $\vec x_1$ at time $t_1$ to end up to position $\vec x_2$ at
time $t_2$. It is convenient for now to consider the (backward) transition probability $Q(\vec x,t)=P(0,0|\vec x,-t)$
that a particle starting at position $\vec x$ at time $-t \leq 0$, ends up at the origin at time $0$ 
(the forward is considered later). 
The latter obeys
the following random backward Kolmogorov equation
\be \label{Eq:BackwardKolmo}
\partial_t Q = D \nabla^2_x Q + \vec f \cdot \vec \nabla_x Q + \vec \xi \cdot \vec \nabla_x Q ,
\ee
with final condition $Q(\vec x,t=0) = \delta^{(d)}(\vec x)$. For simplicity we focus on
$\vec \xi(\vec x,t)$ 
being 
a space-time Gaussian white noise (interpreting (\ref{Eq:BackwardKolmo}) in the \^Ito sense) with variance 
\bea \label{Eq:Defr0}
\overline{\xi_i(\vec x,t)  \xi_j(\vec x',t')} = D r_0^d \delta^{(d)}(\vec x-\vec x') \delta(t-t') \delta_{ij} \ssp ,
\eea  
where the parameter $r_0$ has dimension of a length. Our results on the large scale properties 
should hold for more general distribution of the disorder, as long as correlations of $\vec \xi(\vec x,t)$ are short-ranged in 
space and time. Eq.\eqref{Eq:Defr0} 
can be thought of as an approximation of more realistic models. One is 
a continuum model with 
disorder of 
(dimensionless) magnitude $\sigma_{\xi}$, a finite correlation length $r_c$, 
and a finite correlation time $\tau_c$. In that case $\overline{\xi_i(\vec x,t)  \xi_j(\vec x',t')} =\frac{D}{\tau_c} \sigma_{\xi}^2 R_1(\frac{\vec x -\vec x'}{r_c})  R_2(\frac{t-t'}{\tau_c}) \delta_{ij} $ with $R_1$ and $R_2$ two dimensionless rapidly decaying functions, and one relates $r_0$ to the space-time correlation volume of the noise as $r_0^d \sim \sigma_{\xi}^2 r_c^d  \int d^d \vec y ds R_1(\vec y) R_2(s)$. Another 
is to see the continuum model as 
a limit of a discrete model, and to 
interpret $r_0$ 
as the lattice spacing.

 
In the following we analyze this RW {\em locally around a given
space-time direction} (moving frame velocity) $\vec u \in \mathbb{R}^d$, 
i.e. for $\vec x = \vec u t + \vec{x}'$ with $\vec{x}'=o(t)$. This is equivalent to looking around the origin $\vec{x} = o(t)$
in the model with an effective bias $\vec f_{\vec u}=\vec f + \vec u$: using the equality in law between white noises $\vec{\xi}(t  \vec u + \vec x' , t) \simlaw \vec{\xi}'(\vec x',t)$ one gets $Q_{\vec f}(\vec u t + \vec x',t) \simlaw Q_{\vec f+\vec u}(\vec x',t)$. 
We drop the subscript $\vec u$ in $\vec f_{\vec u}$ unless needed, but $\vec{f}$ should thus be thought of as a control parameter analogous to the velocity of the frame of observation compared to the mean velocity of the particles.
 We first note that the averaged value of the transition probability is equal to the transition probability of a RW in the averaged environment\footnote{This is due to the delta correlations in time in \eqref{Eq:Defr0} and to the \^Ito prescription. As a consequence the bare values of
$D$ and $f$ (or $v$) are not renormalized. A small but finite $\tau_c$ leads to small corrections to
these values.}, hence it is Gaussian and given by $\overline{Q(\vec x,t)}= \frac{1}{ (4 \pi D t)^{\frac{d}{2}}} e^{- \frac{|\vec x + t \vec f |^2}{4 Dt}}$. The regime $|\vec x |= o(t)$ is thus characterized by an exponential decay of the averaged probability: $- \frac{|\vec x + t \vec f |^2}{4 Dt} \simeq  - \frac{\vec f \cdot \vec x}{2D}  - \frac{t |\vec f|^2}{4D}$, hence corresponds to
a large deviation regime, far away from the bulk of the probability, i.e. the optimal direction of the RW $\vec x= -\vec f t$.
To study the local fluctuations around this average profile of the probability, we introduce
the {\it partition-sum} $Z(\vec x,t)$ and {\it height} $h(\vec x,t)$ as 
\be \label{Eq:DefZ}
  Z(\vec x,t) :=  e^{ \frac{\vec f \cdot \vec x}{2D} + t \frac{\vec f^2}{4D} } Q(\vec x,t)   \quad , \quad h(\vec x,t) : = \ln Z(\vec x,t) \ssp .
\ee
Inserting (\ref{Eq:DefZ}) in (\ref{Eq:BackwardKolmo}) we obtain 
\bea \label{Eq:SHE1}
&& \partial_t Z =  D \nabla_x^2 Z + \xi_{{\rm DP}} Z + \vec \xi \cdot \vec \nabla_x Z \ssp , \\
&& \partial_t h =  D \nabla^2 h + D (\vec \nabla  h)^2  +  \xi_{\rm DP} + \vec \xi \cdot \vec \nabla h  \ssp . \label{Eq:KPZ1}
\eea
with the `droplet' initial condition $Z(\vec x,0) = \delta(\vec x)$. In (\ref{Eq:SHE1}), \eqref{Eq:KPZ1} we have introduced the `directed polymer (DP) noise term'
\bea \label{Eq:DPNoiseDef}
\xi_{{\rm DP}} (\vec x,t) = - \frac{\vec f \cdot \vec \xi(\vec x,t)}{2D} \ssp .
\eea
It is a Gaussian white noise with $\langle \xi_{{\rm DP}}(\vec x,t) \xi_{{\rm DP}}(\vec x',t') \rangle = \sigma_{{\rm DP}}^2 \delta(t-t') \delta^{(d)}(\vec x- \vec x')$ and, noting $f := | \vec f|$ the norm of the bias,
\bea \label{Eq:DPNoiseStrength}
\sigma_{{\rm DP}}^2 = \frac{r_0^d }{4 D}  f^2 \ssp .
\eea 
The equations \eqref{Eq:SHE1}-\eqref{Eq:KPZ1}, contain two (mutually correlated) noises. 
While the second source of noise (last term) is a signature of the RW nature of the problem (it is already present in the original backward Kolmogorov equation (\ref{Eq:BackwardKolmo})), the first was generated by our rescaling of the transition probability \eqref{Eq:DefZ} and is a signature of the fact that we
are looking at the large deviation regime: it is the only term in \eqref{Eq:SHE1}-\eqref{Eq:KPZ1}
that depends on $f$.
A crucial observation is that if, in a first stage (justified below), 
one neglects the second source of noise, 
the equations \eqref{Eq:SHE1} and \eqref{Eq:KPZ1} become respectively the multiplicative stochastic-heat-equation (MSHE)
and the KPZ equation \eqref{KPZ1}. 
The solution of the MSHE is known to be the partition sum $Z_{{\rm DP}}(\vec x,t)$ 
of the continuum directed polymer problem, i.e. the equilibrium statistical mechanics at temperature $T=2D$ of directed paths of length $t$, $\vec{x}: \tau \in [0,t] \to \vec x(\tau) \in \JR^d$ with fixed endpoints $\vec x(0)=0$ and $\vec x(t)=\vec x$ in a quenched random potential $-2 D\xi_{{\rm DP}} (t',\vec x(t'))$. It can formally be written as a path-integral
\be
Z_{{\rm DP}}(\vec x,t) = \int_{x(0)=0}^{x(t)=x} {\cal  D} [x] e^{-\frac{1}{2D} \int_{0}^t d\tau  \{ \frac{1}{2} (\frac{d \vec x}{d\tau})^2 -2D \xi_{{\rm DP}} (\tau,\vec x(\tau))  \} } \ssp .
\ee
while the solution of the KPZ equation with the so-called droplet initial condition is given by $h_{{\rm KPZ}}(\vec x,t)=\ln Z_{{\rm DP}}(\vec x,t)$, the two problems hence being, as is well known, equivalent.

The emergence of the MSHE and KPZ equations in this problem is at the core of the 
connection between TD-RWRE and the KPZ universality-class (KPZUC). The rescaling (\ref{Eq:DefZ}) takes into account 
the deterministic influence of the bias $\vec f$ and 
\eqref{Eq:SHE1} sheds light on the peculiar nature of the bias-induced {\it fluctuations} of the transition probability. 
 In the following we argue that these fluctuations dominate the statistical properties of $Z$ at large scale, 
 hence those of $Q$, and that is the mechanism that is responsible for the emergence of KPZ type-fluctuations.
 
 Let us now explore some consequences of this connection in the DP language, which is more adapted
 to the physics of the RW problem in terms of space-time paths. 
 It is well known \cite{kardar1987scaling,Monthus} that the DP exhibits a phase transition as a function of the noise strength $\sigma_{{\rm DP}}$ between: (i) a diffusive phase at small $\sigma_{{\rm DP}} < \sigma_c $ where polymer paths are diffusive $x(\tau) \sim \tau^{1/2}$ and do not feel the influence of the disorder; (ii) a pinned phase at large $\sigma_{{\rm DP}} > \sigma_c $ where directed polymer paths are superdiffusive $x(\tau) \sim \tau^{\zeta_d}$ with $\zeta_d>1/2$ the universal (dimension-dependent) roughness exponent. While in the diffusive phase the fluctuations of the DP free-energy are small, $\ln Z_{{\rm DP}}(t) \sim O(1)$, in the pinned phase the DP optimizes its energy: the partition sum is concentrated on a few optimal paths and the fluctuations of the DP free-energy scale with the length as $\ln Z_{{\rm DP}}(t) \sim t^{\theta_d}$ with $\theta_d = -1 + 2 \zeta_d>0$. While for $d>2$ there is a transition at a non-trivial value $\sigma_c>0$\footnote{The existence of an upper-crical dimension $d_c$ where $\sigma_c= +\infty$ has not yet been settled}, $\sigma_c=0$ in $d=1,2$ and the system is always in the pinned phase. 
 
We now argue, using the interface language, that the second source of noise in \eqref{Eq:SHE1}-\eqref{Eq:KPZ1} is always {\it irrelevant} in the pinned phase at large time. In this phase the KPZ field displays scale invariant fluctuations
and we can rescale $h(\vec x,t) = b^{\alpha} \tilde h(\vec x/b, t/b^z)$ with $b$ large and $z=1/\zeta_d$ and $\alpha=\theta_d/\zeta_d$ the dynamic and roughness exponent of the KPZUC, with $\tilde h=O(1)$. From the scale invariance of the Gaussian white noise, under rescaling the second source of noise in (\ref{Eq:KPZ1}) receives an additional
factor $b^{\alpha-1}$ as compared to the first one. 
This heuristic suggests that the second source of noise is irrelevant as long as $\alpha <1$. This condition is always satisfied
in the rough phase, with $\alpha=1/2$ in $d=1$ and $\alpha$ decreases with $d$.

 This 
leads us to the following conjecture. In the RW problem, looking locally\footnote{Note that in a sense the conservation of probability of the random walk problem
seems to be lost in the KPZ regime. This is only because the mapping to KPZ only holds locally
in the large deviation region $x=o(t)$. Everywhere in that region the probability mass escapes
towards the most probable direction, where the equivalence to KPZ breaks down.}
in the large deviation region $\vec x =o(t)$,
the system undergoes {\it a phase transition as a function of the bias} from: (i) a diffusive phase for $f < f_c$ where 
the local fluctuations of $\ln Q(\vec x,t)$ are $O(1)$ and the random walk paths are diffusive with the same
law as the RW in an averaged environment (for $f=0$ this was shown rigorously in \cite{math1});
 (ii) a pinned phase for $f > f_c$ where $\ln Q(\vec x,t)$ has larger fluctuations scaling as $t^{\theta_d}$
 and random walk paths are superdiffusive with the DP roughness exponent $\zeta_d$. In addition 
the full multi-point distribution of $\ln Q(\vec x,t)$ at large $t$ are expected to be universal 
and identical to those of the free-energy $\ln Z_{{\rm DP}}(\vec x , t)$ of the DP problem in the pinned phase. 
Furthermore $f_c=0$ in $d=1,2$ and in $d > 2$ we can give an estimate of the transition point.
For the KPZ equation \eqref{KPZ1}, $d=2+\epsilon$ renormalization group (RG) calculations indicate that the transition for $d>2$ occurs for the dimensionless coupling\footnote{$K_d=S_d/(2 \pi)^d$ where $S_d$ is the $d$-dimensional unit sphere area.} $g:=K_d \Lambda^{d-2} \frac{\lambda_0^2 D_0}{8 \nu_0^3} = g_c$ of order $\epsilon$: $g_c=\epsilon+ O(\epsilon^2)$, with $\Lambda^{-1}$ a short distance cutoff \cite{KPZ,FreyTauber}. Translating into the RW with $\Lambda=1/r_c$ we
find $g_c=K_d \sigma_\xi^2 r_c^2  f_c^2/(8 D^2)$, which provides an estimate for $f_c$. As we mentionned the bias also incorporates the effect of looking at the problem in a moving frame of velocity $\vec{u}$. The phase transtion can thus be driven by $\vec{u}$ and occurs when $|\vec{f}_{\vec{u}}| = |\vec{f} + \vec{u}|= f_c$: the pinned phase occurs everywhere in space outside a `light-cone' around the optimal direction of the RW (see Fig.~\ref{fig:NicePic}).
This picture agrees with known results. It was indeed shown in \cite{yilmaz1,yilmaz2,yilmaz3} that the annealed and quenched large deviations rate functions of an unbiased lattice RW, respectively defined as $I_{{\rm a}}(|\vec u|):= - \lim_{t \to \infty} \frac{\ln \overline{Q(t \vec u,t)}}{t}$, and $I_{{\rm q}}(|\vec u|):= - \lim_{t \to \infty} \frac{\overline{\ln Q(t \vec u,t)}}{t}$ satisfy the following properties: (i) $I_{{\rm a}}(0) = I_{{\rm q}}(0) =0 $ (optimal direction); (ii) $I_{{\rm a}}(u) < I_{{\rm q}}(u) $ $\forall u \neq 0$ in $d=1,2$ and for $u$ large enough in $d\geq 3$; (iii) $I_{{\rm a}}(u) = I_{{\rm q}}(u)$ for $u$ small enough in $d \geq 3$. This confirms our scenario of a transition in $d \geq 3$, and our arguments show that the strong bias phase
 is in the KPZ class. 

\begin{figure}
\centerline{\includegraphics[width=8.5cm]{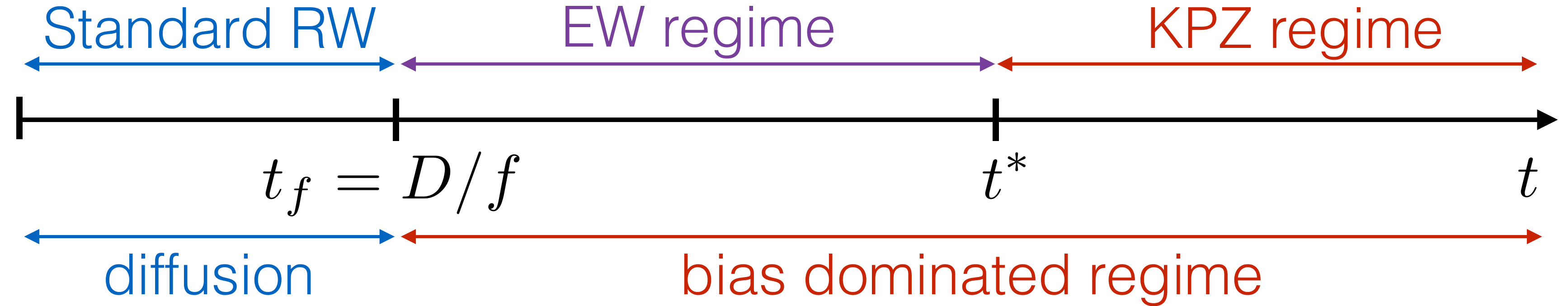}} 
\caption{The behavior of the RW crosses over from a diffusion dominated regime to a bias dominated regime when $t\sim t_f$. This second regime can also be subdivided in between a EW regime at small times and a KPZ regime  if $t^* \gg t_f$ (see text for an estimation of $t^*$ in $d=1,2$).
}
\label{fig:CrossoverPic}
\end{figure}

\begin{figure}
\centerline{\includegraphics[width=8cm]{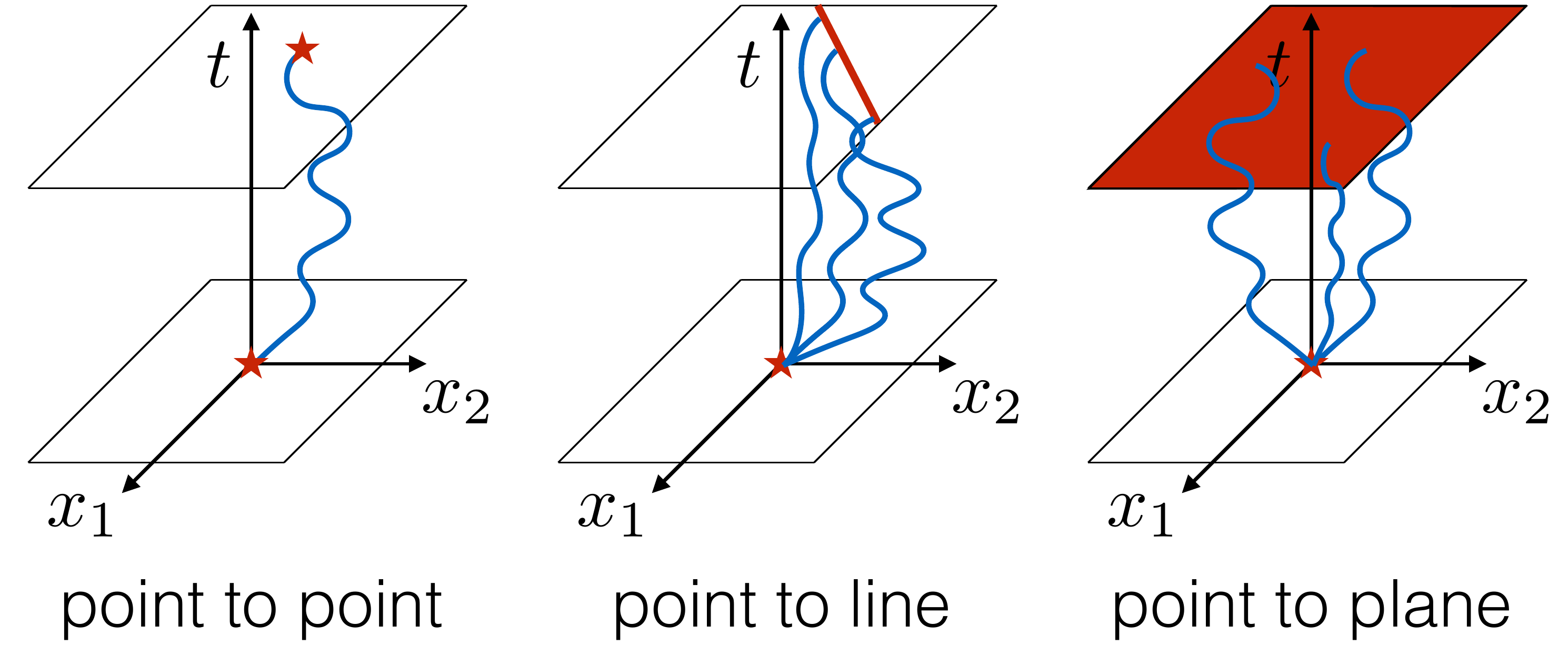}} 
\caption{Some typical polymer geometries for DP in $d=2$. See text for applications to the RW problem.}
\label{fig:GeometriesPic}
\end{figure}

\medskip

{\bf Scales and crossovers} Let us now discuss the scale at which KPZUC emerges, first in the simpler one-dimensional case.
To that aim, note that rescaling time, space and height in \eqref{Eq:KPZ1}
as $t = t^* t'$, $x = x^* x'$ and $h'(t',x') := \frac{1}{h^*} h(t^* t' , x^* x')$ with the characteristic scales $t^* = \frac{(4 D)^3}{r_0^2 f^4}$,
$x^* = \frac{8 D^2}{r_0 f^2} $ and $h^*=1$, leads to a rescaled KPZ-like equation for $h'(t',x') $ identical to \eqref{KPZ1}
with $\lambda_0=D_0=2$, $\nu=1$, up to the second source of noise of 
(\ref{Eq:KPZ1}) which now involves a unit white noise multiplied by the dimensionless ratio $f r_0/(2 D\sqrt{2})$.
Hence for $fr_0/D \ll 1$ (weak-bias/weak-noise or large diffusivity limit) the `deformed' KPZ-equation (\ref{Eq:KPZ1}) becomes equivalent to the standard KPZ equation 
(this is reminiscent of the `weak-universality' of the KPZ equation). Hence in this weak bias regime,
we can apply the known results for the continuum KPZ equation, see \cite{SM}. Thus, for $t/t^*\gg1$, we 
predict that the KPZUC appears in the RW problem. At short scale $t',x' \ll 1$, the behavior of the height in the KPZ equation becomes similar
to the Edward-Wilkinson (EW) behavior \cite{gueudre2012short}. In the RW problem we expect by inspection of (\ref{Eq:KPZ1}) that the first source of noise (bias) dominates for $x \gg  x_f = D/f$ while the second (diffusion) dominates for $x \ll x_f$ (with an associated time-scale $t_f:=x_f^2/D$). We conclude that for $fr_0/D \ll 1$ there is a regime $x_f \ll x \ll  x^*$ and $t_f \ll t \ll t^*$ where one can already neglect the second source of noise but KPZUC type fluctuations have not yet been build up: this should be an EW regime\footnote{We note that links between the Edward-Wilkinson universality class and the TD-RWRE have already been studied, see \cite{RAP,RAP2}. This however seems very different from what we discuss here.}, see Fig.~\ref{fig:CrossoverPic}.

In general $d$ the scale at which the bias starts to dominate remains $t_f$ and $x_f$, but the scales $t^*$ and $x^*$ where KPZ fluctuations emerge change. For example in $d=2$ disorder is marginally relevant and from 
RG \cite{KPZ,FreyTauber,NelsonPLD} one has $x^* \simeq_{g \ll 1} \Lambda^{-1} e^{1/g}$ with 
here (see above) $g=r_0^2 f^2/(16 \pi D^2)$, $t^*=(x^*)^2/D$ and for the RW we take $\Lambda^{-1} = r_c$. 
For $g \ll 1$ the scales are well-separated and we similarly expect an intermediate EW regime of fluctuations.


{\bf Discrete models} To show the versatility of our argument, we now consider discrete models of TD-RWRE on $\mathbb{Z}^d$. We note\footnote{Here and in the following the index ${\sf d}$ emphasizes the discrete nature of the coordinate.}
$\vec x_{\sd}(t_{\sd}) \in \mathbb{Z}^d$ the position of the random walker at time $t_{\sd} \in \JN$. At $t_{\sf d}$, the particle jumps, $\vec x_{\sd}(t_{\sd}+1) - \vec x_{\sd}(t_{\sd}) =\epsilon_i \vec e_i$, with probability $p_{t_{\sd},\vec x_{\sd}}^{\epsilon_i \vec e_i}$. 
Here $\{ \vec e_{i}  , i = 1, \cdots,  d \}$ is the set of unit translation vectors on the lattice and $\epsilon_i = \pm 1$ 
gives the jump direction. The $p_{t_{\sd},\vec x_{\sd}}^{\epsilon_i \vec e_i}$ are independent (except if they leave from the same site, where $ \sum_{i=1}^d \sum_{\epsilon_i = \pm 1} p_{t_{\sd} , \vec x_{\sd}}^{\epsilon_i \vec e_i}  = 1$) and identically distributed (iid). We suppose that these are biased: $\overline{p^{\epsilon_i \vec e_i}} = 1/(2d) + f_{\sd}^{\epsilon_i \vec e_i}$ with at least some of the $f_{\sd}^{\epsilon_i \vec e_i}$ different from $0$, and introduce the centered random variables $\xi_{t_{\sd},\vec x_\sd}^{\epsilon_i \vec e_i} = p_{t_{\sd} , \vec x_\sd}^{\epsilon_i \vec e_i} - \overline{p_{t_{\sd} , \vec x_\sd}^{\epsilon_i \vec e_i}}$. 
As in the continuum we consider $Q_{t_{\sd}}(\vec x_{\sd})$, the probability that a particle starting at $\vec x_{\sd}$ at time $-t_{\sd}$ hits the origin at time $0$. Introducing a lattice spacing $a$, we now estimate the KPZ noise that appears in this discrete RW model when rescaled diffusively around the diagonal of the lattice with $t_{\sd} \simeq D t/a^2 $ and $\vec x_{\sd} \simeq  \Lambda^{-1} \vec{x}/a$ ($\Lambda$ is a diagonal matrix), rescaling also the discrete noise as $\xi^{\epsilon_i \vec{e}_i} \sim a^{y_d} \hat{\xi}^{\epsilon_i \vec{e}_i} $. In \cite{SM} we show {\it along the same line than in the continuum} that the partition sum variable
\bea
Z(\vec x,t)  \simeq_{a \to 0} A^{-t_{\sd}} \prod_{i=1}^d B_i^{-x_i} Q_{t_{\sd}}(\vec x_{\sd}) \ssp ,
\eea
with $A := 2 \sum_{i=1}^d  \sqrt{\overline{p^{-\vec e_i}}  \cdot \overline{p^{\vec e_i}}}$, $B_i:=\sqrt{  \overline{p^{-\vec e_i}} / \overline{p^{\vec e_i}}}$ and $\Lambda = {\rm diag}(\sqrt{A/ \sqrt{\overline{p^{\vec e_i}} \cdot \overline{p^{-\vec e_i}}  }})$ satisfies the MSHE $\partial_{t} Z = D \nabla^2 Z + \xi_{{\rm DP}}(\vec x , t)$ with $\xi_{{\rm DP}}(\vec x , t)$ a Gaussian white noise of strength $\sigma_{{\rm DP}}$ given by
\bea \label{Eq:DPNoiseStrength2}
\sigma_{{\rm DP}}^2=   C a^{d-2+2y_d}  \overline{\left(\sum_{i=1}^d \sum_{\epsilon_i = \pm 1}  (B_i^{\epsilon_i} -1) \hat{\xi}_{\frac{a^2}{D_d} t,a \Lambda \vec x}^{\epsilon_i \vec e_i}  \right)^2} \ssp ,
\eea
with $C := \frac{ 2^{d} D {\rm det}(\Lambda )}{A^2 }$. Note that the lattice spacing $a$ still appears in this expression. In $d=1$ the result reads $\sigma_{{\rm DP}}^2 = \frac{32 \sqrt{2} D f_{\sd}^2}{(1-4f_{\sd}^2)^2}  a^{-1 + 2 y_1} \overline{\hat{\xi}^2}$ and one can obtain a finite limit as $a \to 0$ by taking $y_1 = 1/2$: this is a weak-noise limit and there the convergence to the MSHE is an exact result (similar but different from \cite{CorwinRWREWeakU}) that we check in \cite{SM} using the Beta polymer. To compare with the continuum one should also rescale the discrete bias with the lattice spacing by taking $f_{\sd} \sim a$ (see \cite{SM}) and keeping $\xi \sim 1$. In that case one obtains a small noise $\sigma_{{\rm DP}}^2 \sim a$, in complete analogy with the continuum in $d=1$ (\ref{Eq:DPNoiseDef}) if one takes $a \equiv r_0$. In general (\ref{Eq:DPNoiseStrength2}) should be considered as an estimate of the DP noise felt in this discrete setting that, in the scaling $f_{\sd} \sim a$, coincides with the continuum result (\ref{Eq:DPNoiseDef}), but also generalizes it in the large bias regime $f_{\sd} = O(1)$ where the continuum model does not apply anymore.

\medskip

{\bf Universal distributions} 
It is useful to extend our analysis to the {\it forward} transition probability $P(\vec{x},t) = P(\vec{x},t | 0,0)$. It satisfies the Fokker-Planck equation $\partial_t P = D \nabla_x^2 P - \vec{\nabla}_x \cdot ((\vec f + \vec{\xi})P)$. Considering again the `partition sum variable' $Z(\vec x , t) := e^{-\frac{\vec f \cdot \vec x}{2D} + t \frac{f^2}{4D}} P(\vec x , t) $ generates additional noise terms in this equation and our arguments can be repeated (see \cite{SM}): the statistical properties of $Z(\vec x , t)$ at large scale are identical to those of the DP partition sum. In fact note that in law we must have $P(\vec{x},t) \sim Q(-\vec{x},t)$.

We can also consider different initial/final conditions in the forward/backward setting. This is of great interest since the KPZUC is splitted in sub-universality classes \cite{CorwinReviews} that depend on the boundary conditions, and we thus predict universal distributions for the fluctuations of $\ln P$ or $\ln Q$ according to the chosen boundary conditions. These were determined numerically in $d=2$ \cite{halpin2012-2D} and are known analytically in $d=1$, on which we now focus. Using our argument and KPZ universality, we conjecture that the appropriately rescaled fluctuations of $\ln P(x,t)$ and $\ln Q(x,t)$ are universal 
in the large-deviation region and distributed as a TW GUE random variable $\chi_2$ \cite{CorwinReviews}. 
This has already been observed analytically and numerically for the exactly solvable Beta polymer, see \cite{BarraquandCorwinBeta,usBeta}. For the continuum model \eqref{langevin}-\eqref{Eq:Defr0}
in the absence of bias, $f=0$, but in a moving frame, we obtain (using \cite{droplet}, see \cite{SM}) a sharp prediction for $t \gg t^*$
\be \label{Eq:ResDropletLetter}
\ln P(x=u t,t)  \simeq - I_{\rm q}(u) t + \lambda(u) t^{1/3} \chi_2
\ee
where $I_{\rm q}(u) \simeq \frac{u^2}{4 D} 
+ \frac{2 r_0^2 u^4}{3 (8 D)^3}$
and $\lambda(u) \simeq 
\frac{r_0^{2/3} u^{4/3}}{4D}$,
estimates valid 
in the weak bias limit $r_0 u/D \ll1$,
i.e. $t^*=(4 D)^3/(r_0^2 u^4) \gg t_u=D/u^2$. These arguments
extend \cite{SM} using the KPZ equation at finite $t/t^*$. They indicate
an intermediate regime $x \sim t^{3/4}$ between typical Gaussian diffusion $x \sim (D t)^{1/2}$
and the large deviation $x \sim t$ regime. Defining the dimensionless
variable $y$ through $x=y r_0(4 D t/r_0^2)^{3/4}$, a crossover
from EW to KPZ fluctuations in \eqref{Eq:ResDropletLetter} occurs as $t/t^*=y^4=O(1)$
increases. The crossover to diffusion occurs for $x \sim (D t)^{1/2}$ and 
we predict on that scale fluctuations $\sim (r_0^2/(D t))^{1/4}$: fluctuations decay with time and we retrieve that $P(x,t)$ converges (almost surely) to $\overline{P(x,t)}$ in this regime.

\begin{figure}
\centerline{\includegraphics[width=7.5cm]{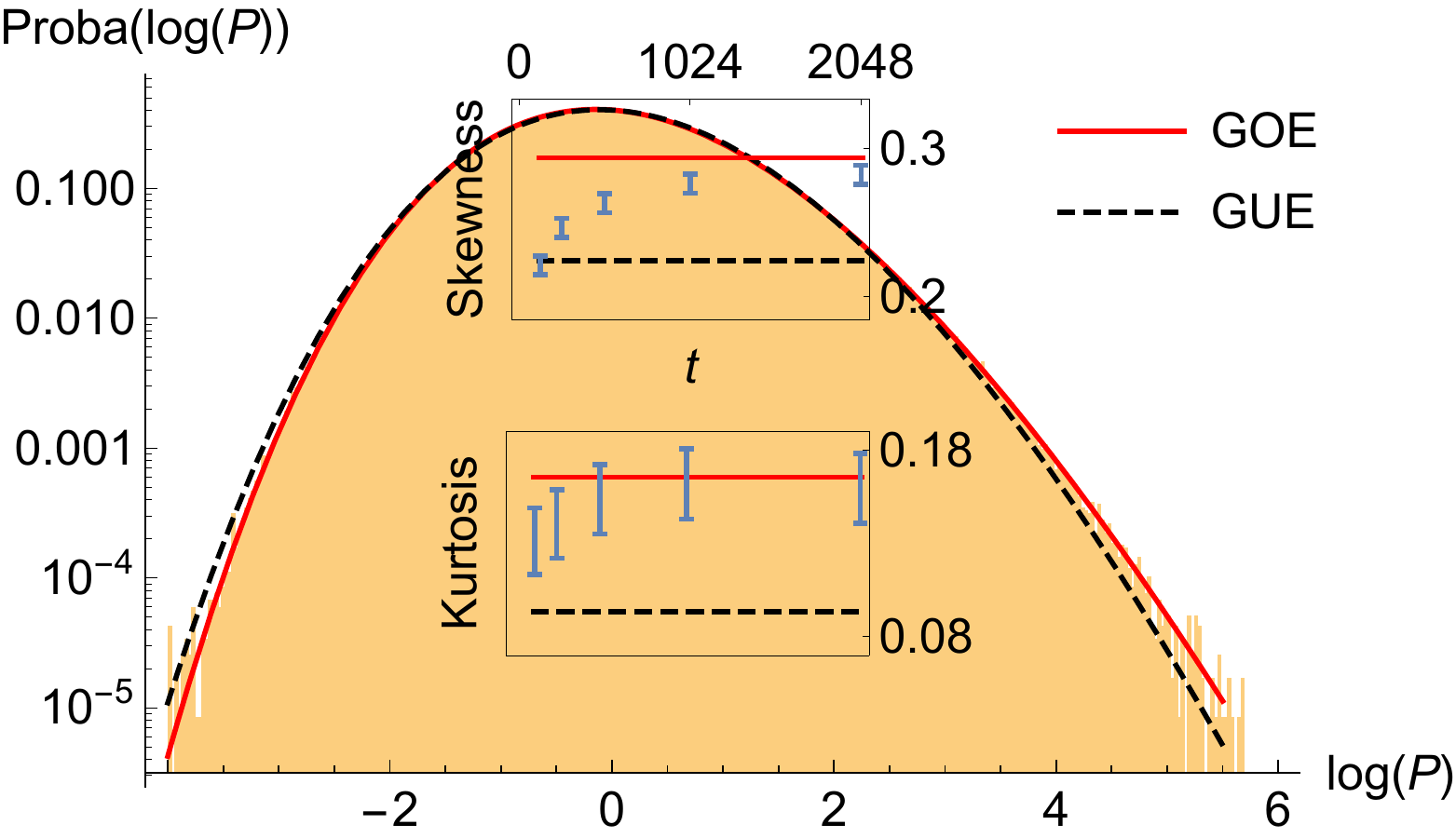}} 
\caption{Numerical observation of the GOE TW distribution in the large time fluctuations of the log of the forward transition probability $P(0,t_{\sd})$ of a TD-RWRE on $[-4096,4096]\cap\JZ$ with reflexive boundary conditions in a biased random environment (see details in \cite{SM}), starting at time $t_{\sd}=0$ with $P(x_{\sd} , t_{\sd}=0)$ given by the stationary measure of the RW in the absence of disorder. Main plot: centered and normalized histogram (in a logarithmic scale) of $\ln P(0,t_{\sd})$ with $t_{\sd}=2048$ compared with the GOE (red line) and GUE (black-dashed line) TW distribution. The insets show the convergence of the skewness (top) and of the excess of kurtosis (bottom) of the distribution of  $\ln P(0,t_{\sd})$ to values close to those of the GOE TW distribution. Error-bars are 3-sigma Gaussian estimates.}
\label{fig:GOE}
\end{figure}

We now make a prediction related to the flat KPZ sub-universality class, which as yet has never been
observed in the TD-RWRE context. It is known that the large time fluctuations of the logarithm of the solution of the MSHE $\partial_t Z = D \nabla^2 Z + \xi_{{\rm DP}} Z$ with flat initial condition $Z(x,t=0) = 1$, properly scaled, are distributed according to a GOE Tracy-Widom random variable $\chi_1$. Here it means that we must start the RW with the initial\footnote{Here we adopt the forward setting.
Indeed, observing GOE fluctuations in the backward case requires imposing the final probability 
$Q(x,t=0) \sim e^{-f x/D}$ which does not seem possible.}
probability distribution given by $P(x,t=0) \sim e^{f x/(2 D)}$. While non-normalizable in the infinite space, it is a natural initial condition on an interval of length $L$ with reflecting boundary conditions, $x \in [-L/2 , L/2]$: it is the stationary measure of the RW in the absence of disorder. Turning on the disorder at $t=0$ we predict that at large time (in the regime $1 \ll t/t^* \ll (L/x^*)^{3/2}$ to avoid the influence of the boundaries), $\ln P(0,t)$ fluctuates as $c (t/t^*)^{1/3} \chi_1$, where
$c=2^{-2/3}$ \cite{we-flat} when $t^* \gg D/f^2$. This scenario, and its universality, is checked explicitly through simulations of a $1-$dimensional discrete TD-RWRE, see Fig.\ref{fig:GOE}.

\medskip

{\bf Extreme value statistics} An important application of the large deviation regime of the RW where the KPZUC emerges, is to
extreme value statistics. Consider
$N \gg 1$ independent walkers starting at the origin at $t=0$ with no bias, $\vec f=0$.
We define $x_{{\rm max}}(t) := {\rm max}_{i=1,\cdots,N} \{\vec x_i(t) \cdot \vec{e}_1 \}$ the position of the rightmost walker in the direction of the unit vector $\vec{e}_1$. We show \cite{SM} that the 
KPZ-universality in the fluctuations of the logarithm of the transition probability, $\ln P(\vec x \cdot \vec e_1 = u t,t|0,0)$, implies that as 
$N,t \to \infty$ 
with 
\be
\gamma= \frac{\ln N}{t}  \quad \text{fixed} \, ,
\ee
then $x_{\rm max}(t)$ grows ballistically, with equality in law
\bea \label{Eq:Xmax}
x_{{\rm max}}(t) \simeq u^*_\gamma t  + c(\gamma) t^{\theta_d} \chi + o(t^{\theta_d}) \ssp .
\eea
Here $\theta_d$ is the KPZ exponent, $\chi$ is a universal distribution \cite{SM} characteristic of the point to hyperplane (of dimension $d-1$) subuniversality KPZUC (see
e.g. \cite{halpin2012-2D}). Here $c(\gamma)$ and $u^*_\gamma$ are non-universal, 
given in the continuum in \cite{SM}. 
This is valid if the front 
velocity $u^*_\gamma>u_c$, 
so that KPZUC appears 
(with $u_c=0$ in $d=1,2$). A formula such as (\ref{Eq:Xmax}) was rigorously shown in an exactly solvable 1D 
model
in \cite{BarraquandCorwinBeta} 
with $\theta_{d=1}=1/3$ and $\chi =\chi_2$ 
a GUE TW random variable. Similarly,
the first arrival time at ${\vec x} \cdot \vec e_1=\ell$, $T_{{\rm Hit}}(\ell)$, of a particle from a cloud of $N$ independent particles, behaves, for fixed $\hat \gamma= \frac{\ln N}{\ell}$ as
\bea
&& T_{{\rm Hit}}(\ell) \simeq \ell/v_{\hat \gamma}^* - d(\hat \gamma) \ell^{\theta_d} \chi + o(\ell^{\theta_d}) 
\eea
with the same universal random variable $\chi$ \cite{SM}. Arrival times in compact domains, i.e. a ball, leads instead to point to
point KPZ distribution in any $d$.

\medskip

{\bf Conclusion} In this Letter we investigated the origin and consequences of the emergence of universal statistics of the KPZUC in the large deviations regime of TD-RWRE in arbitrary dimension. We focused on short range correlated random media but our method
readily extends to long range (LR) {\it spatial} correlations \cite{SM}, leading to the distinct LR space correlated KPZ universality
classes \cite{KardarLR}. Important questions for the future are how LR correlations in {\it time} in the medium, 
and interactions within a cloud of $N$ particles, will affect the results, since those are present
in many natural examples, such as the atmosphere or the ocean. We hope that this motivates further connections 
between the fields of growth and diffusion.

\medskip

{\it Acknowledgments} We thank G. Barraquand, I. Corwin and F. Rassoul-Agha for discussions. T.T. has been supported by the InterUniversity Attraction Pole phase VII/18 dynamics, geometry and statistical physics of the Belgian Science Policy. This research was supported in part by the National Science Foundation under Grant No. NSF PHY11-25915 and we acknowledge hospitality from the KITP in Santa Barbara.

\newpage

\begin{widetext}

\begin{center}
{\Large Supplemental Material}
\end{center}

We give here some details on the derivation of the results presented in the main text, as well as a non-trivial test of one of our result.

\bigskip

{\bf The continuum RWRE model as a formal limit of the discrete one}

Here we do a formal calculation showing how the discrete and continuum RWRE model introduced in the letter are related to one another.

We start from the discrete backward Kolmogorov equation satisfied by the transition probability $Q_{t_{\sd}}(\vec x_{\sd})$ in the discrete RWRE model (here we have changed $ p_{t_{\sd} , \vec x_{\sd}}^{\epsilon_i \vec e_i} \to p_{-t_{\sd} , \vec x_{\sd}}^{\epsilon_i \vec e_i}$ compared to the notation of the letter):
\bea \label{Sm:1:Eq1}
&& Q_{t_{\sd}+1}(\vec x_{\sd}) = \sum_{i=1}^d \sum_{\epsilon_i = \pm 1} p_{t_{\sd}+1 , \vec x_{\sd}}^{\epsilon_i \vec e_i} Q_{t_{\sd}}( \vec x_{\sd}+ \epsilon_i \vec e_i)  \\
&& Q_{t_{\sd}=0}(\vec x_{\sd}) = \prod_{i=1}^d \delta_{x_i , 0}  \ssp .  \nn
\eea
We split the hopping probabilities into the averaged and fluctuating parts $p_{t_{\sd} ,\vec{x}_\sd}^{\epsilon_i \vec e_i} = \frac{1}{2d} + f_{\sd}^{\epsilon_i \vec e_i} + \xi_{t_{\sd},\vec x_{\sd}}^{\epsilon_i \vec e_i}$ with $\sum_{i=1}^{d} \sum_{\epsilon_i = \pm1} f_{\sd}^{\epsilon_i \vec e_i}  =\sum_{i=1}^{d} \sum_{\epsilon_i =\pm1} \xi_{t_{\sd},\vec x_\sd}^{\epsilon_i \vec e_i} = 0$. Introducing a lattice spacing $a>0$ we take
\bea \label{SM:EQ:Rescalingfd}
f_{d}^{ \epsilon_i \vec{e}_i} = a  \frac{1}{2D \sqrt{2d }}\epsilon_i \vec f \cdot \vec e_i \quad , \quad \vec x_{\sd} = \frac{\vec x}{a \sqrt{2d}} \quad , \quad t_{\sd} = \frac{D t}{ a^2}
\eea
and consider the formal limit
\bea
Q(\vec x , t) =  \lim_{2 a \to 0} \frac{1}{2 a} Q_{t_{\sd}}(\vec x_{\sd}) \ssp .
\eea
The $1/2$ factor here and some that will appear below are linked to the fact that according to (\ref{Sm:1:Eq1}), $Q_{t_{\sd}}(\vec x_{\sd})$ is $0$ if the parity of the spatial and temporal coordinates are not equal. We also suppose that $\xi_{t_{\sd}+1,\vec x_{\sd}}^{-\vec e_i} = -\xi_{t_{\sd}+1,\vec x_{\sd}}^{\vec e_i}$ and that $\xi_{t_{\sd}+1,\vec x_{\sd}}^{\vec e_i}$ and $\xi_{t_{\sd}+1,\vec x_{\sd}}^{\vec e_j}$ are iid (for $i \neq j$) to match the continuum. Inserting this scaling into (\ref{Sm:1:Eq1}) and expanding leads to
\bea
Q(\vec x , t) + \frac{a^2}{D} \partial_t Q(\vec x , t) \simeq  Q(\vec x , t) + a^2 \nabla_x^2 Q (\vec x , t) + \frac{a^2}{D} \vec f \cdot \vec \nabla_x  Q (\vec x , t) +  2a \sqrt{2d} \sum_{i=1}^d   \xi_{t_{\sd}+1,\vec x_{\sd}}^{\vec e_i}   \partial_{x_i} Q(\vec x , t)  \ssp .
\eea
We now want to understand the leading behavior of the remaining  noise term in the equation and write a reasonable continuous analogue of it. To this aim we calculate the generating function of $\hat{\vec{\xi}}(\vec x , t) :=  2 a \sqrt{2d} \sum_{i=1}^d \xi_{t_{\sd}+1,\vec x_{\sd}}^{\vec e_i} \vec{e}_i$, introducing an arbitrary test function $\vec \phi(\vec x , t)$: 
\bea
\overline{e^{ \int_{0}^{1}dt \int_{[0,1]^d} d^d x \vec{\hat{\xi}}(\vec x , t) \cdot \vec{\phi}(\vec x , t) }}  && \simeq \overline{e^{  \frac{a^2}{D}\sum_{t_{\sd}=0}^{D/a^2} (2a \sqrt{2d})^d \sum_{ x_1/2= 0 }^{\frac{1}{2a\sqrt{2d}}} \cdots \sum_{x_{d}/2= 0 }^{\frac{1}{2a \sqrt{2d}}} (2 a \sqrt{2d} \sum_{i=1}^d \xi_{t_{\sd}+1,\vec x_{\sd}}^{\vec e_i} \phi_i( \frac{a^2}{D} t_{\sd}, \frac{1}{a} \vec x_{\sd})) }} \nn \\
&&  \simeq e^{   \frac{1}{2} \sum_{t_{\sd}=0}^{D/a^2}  \sum_{x_1/2= 0 }^{\frac{1}{2a\sqrt{2d}}} \cdots \sum_{x_{d}/2= 0 }^{\frac{1}{2a\sqrt{2d}}}   \overline{ \left( \frac{a^2}{D} (2a \sqrt{2d})^d 2a \sqrt{2d}  \sum_{i=1}^d \xi_{t_{\sd}+1,\vec x_{\sd}}^{\vec e_i}  \phi_i(\frac{a^2}{D}  t_{\sd},\frac{1}{a} \vec x_{\sd})  \right)^2}    + \cdots}  \nn \\
&& \simeq e^{  \frac{(2a \sqrt{2d})^{d+2} }{2D}   \int_{0}^{1}dt \int_{[0,1]^d} d^d x   \sum_{i=1}^d  \overline{(\xi_{t_{\sd}+1,\vec x_{\sd}}^{\vec e_i})^2    } (\phi_i(\vec x , t))^2   + \cdots }   \nn \\
&& \simeq e^{ \frac{1}{2}\int_{0}^{1}dt \int_{[0,1]^d} d^d x  \sum_{i=1}^d  \overline{(\hat{\xi}_{i}(\vec x , t))^2  (\phi_i(\vec x , t))^2 }   + \cdots  }\ssp .
\eea
Where here the discrete sums take into account that the relevant discrete noise variables $\xi_{t_x,\vec x_{\sd}}$ are those for which $t_{\sd}$ and the coordinates of $\vec x_{\sd}$ have the same parity. This indicates that to leading order in $a$ the noise $\hat{\vec{\xi}}(\vec x , t)$ can be taken as a Gaussian white noise with $\overline{\hat{\xi}_{i}(\vec x , t) \hat{\xi}_j(t',\vec x') } =  \delta_{ij} \delta(t-t')\delta^{(d)}(\vec x - \vec x')  \frac{1}{D} (2 a \sqrt{2d})^{d+2} \overline{(\xi_{t_{\sd}+1,\vec x_{\sd}}^{\vec e_i})^2 }$. This suggests that $Q(\vec x , t)$ formally satisfies the continuum backward Kolmogorov equation

\bea
\partial_t Q(\vec x , t) \simeq  D \nabla_x^2 Q (\vec x , t) + \vec f \cdot \vec \nabla_x  Q (\vec x , t)  + \vec{\xi}(\vec x , t) \cdot \vec \nabla_x Q(\vec x , t)\ssp ,
\eea
with a Gaussian white noise $\vec{\xi}(\vec x , t)$ with covariance
\bea
\overline{\xi_{i}(\vec x , t) \xi_j(\vec x',t') } =  D (2 \sqrt{2d})^{d+2} a^d \overline{(\xi_{t_{\sd}+1,\vec x_{\sd}}^{\vec e_i})^2   } \delta(t-t')\delta^{(d)}(\vec x- \vec x') \delta_{ij} \ssp .
\eea
Hence taking $\overline{(\xi_{t_{\sd}+1,\vec x_{\sd}}^{\vec e_i})}^2 = \frac{1}{ (2\sqrt{2d } )^{d+2}}$ we formally retrieve the continuum model studied in the letter with $ r_0 \equiv a$. Let us conclude for clarity by a few remarks on the meaning of this non-rigorous calculation.

\medskip

These considerations are obviously formal since the identified continuum noise scales with the lattice spacing $a$, and is therefore strictly $0$ in the continuum limit. This is because taking the continuum RWRE model as a large scale description of the discrete one only makes sense close to the optimal direction (with a disrete bias scaling with the lattice spacing as $f_{\sd} \sim a$) where an uncorrelated noise is in fact irrelavant at large scale, as is hinted by this calculation: we retrieve that $d_c=0$ for space-time uncorrelated noise in the normal regime of fluctuations. The continuum RWRE model with delta-correlated white noise studied in the letter cannot in fact be obtained as a limit of the discrete model. On the other hand, if one spatially smoothes the continuum noise on a scale of order $r_c$, the continuum model can then be obtained as a limit of a discrete model if one starts with a discrete model where the noise is spatially correlated on a scale of order $r_c/a$. 

Overall we want by this calculation to highlight the importance of the correlation length of the noise in the RWRE problem. Note also (as is explored in the letter) that on the lattice a discrete uncorrelated noise can have important effects at large scale if one looks away of the optimal direction with $f_{\sd} = O(1)$. In that case however the continuum and discrete RWRE models cannot be easily compared anymore.

\bigskip

{\bf Estimation of the DP-noise in the discrete model}

Here we justify the result (\ref{Eq:DPNoiseStrength2}) presented in the letter.
We start again from (\ref{Sm:1:Eq1}) with the decomposition of the discrete noise into fluctuating and non fluctuating parts. We perform a rescaling of $Q_{t_{\sd}}(\vec x_{\sd})$ as, introducing a discrete partition sum variable $\check Z_{t_{\sd}}(\vec x_{\sd})$:
\bea \label{Eq:SM:DiscreteRescaling11}
Q_{t_{\sd}}(\vec x_{\sd}) = A^t_{\sd} \left( \prod_{i=1}^d B_i^{x_i}  \right) \check Z_{t_{\sd}}(\vec x_{\sd})  \ssp ,
\eea
with
\bea
&& (  \frac{1}{2d} + f_{\sd}^{\vec e_i} ) B_i = (  \frac{1}{2d} + f_{\sd}^{-\vec e_i} )/B_i = \tilde{B}_i  \nn \\
&&  B_i = \sqrt{ \frac{  \frac{1}{2d} + f_{\sd}^{-\vec e_i}}{  \frac{1}{2d}+ f_{\sd}^{\vec e_i}}} \quad , \quad \tilde{B}_i = \sqrt{(  \frac{1}{2d} + f_{\sd}^{-\vec e_i})(  \frac{1}{2d} + f_{\sd}^{\vec e_i}))} \ssp .
\eea
The remaining parameter $A$ is fixed below. We assume the existence of the following continuum diffusive limit:
\bea \label{Eq:SM:DiscreteRescaling12}
Z_{t} (\vec{x}) = \lim_{a \to 0} \check Z_{t_{\sd} = D\frac{t}{a^2}} (\vec x_{\sd} = \frac{\Lambda^{-1}  \vec x}{a} )
\eea
where $\Lambda$ is a diagonal matrix fixed below, $D$ is a constant and $Z_{t} (\vec{x})$ is sufficiently smooth to permits the expansion below. Inserting everything in (\ref{Sm:1:Eq1}) we obtain:
\bea
&& Z  + \frac{a^2}{D} \partial_{t} Z  + o(a^2) = \frac{1}{A}\sum_{i=1}^d \tilde{B}_i  \left( 2 Z+    \Lambda_i^2 a^2 \partial_{x_i}^2 Z  \right)  + \frac{1}{A} \sum_{i=1}^d \sum_{\epsilon_i = \pm 1}  (B_i^{\epsilon_i} -1) \xi_{\frac{a^2}{D_d} t,a \Lambda \vec x}^{\epsilon_i \vec e_i}    Z  \\
&& +\frac{a}{A} \sum_{\epsilon_i = \pm 1}  B_i^{\epsilon_i} \xi_{\frac{a^2}{D_d} t,a \Lambda \vec x}^{\epsilon_i \vec e_i} \partial_{x_i} Z + \cdots \nn
\eea
where we have used the conservation of probability $\sum_{i=1}^d \sum_{\epsilon_i = \pm 1}=0$ to write $ \sum_{i=1}^d \sum_{\epsilon_i = \pm 1}  B_i^{\epsilon_i} \xi_{\frac{a^2}{D_d} t,a \Lambda \vec x}^{\epsilon_i \vec e_i}    Z = \sum_{i=1}^d \sum_{\epsilon_i = \pm 1}  (B_i^{\epsilon_i} -1) \xi_{\frac{a^2}{D_d} t,a \Lambda \vec x}^{\epsilon_i \vec e_i}    Z $. In order for the deterministic term on the right-hand side to give the istropic Laplacian on $\JR^d$ we thus choose as in the letter
\bea
A = 2 \sum_{i=1}^d \tilde{B}_i  \quad , \quad \Lambda_i = \sqrt{A/\tilde{B_i}}
\eea
and we obtain
\bea \label{SM:EQ:Derivation2}
&& \partial_{t} Z = D \nabla_x^2 Z + \frac{D}{A a^2} \sum_{i=1}^d \sum_{\epsilon_i = \pm 1}  (B_i^{\epsilon_i} -1) \xi_{\frac{a^2}{D_d} t,a \Lambda \vec x}^{\epsilon_i \vec e_i}    Z   + \cdots 
\eea
Where the $\cdots$ above contain additional noise terms that are subdominant compared to one that has been retained, which naively appear of order $O(1/a^2)$. We now investigate what is the appropriate continuum limit of this noise, eventually rescaling the discrete noise as
\bea
\xi_{t_{\sd} , \vec{x}_{\sd}}^{\epsilon_i \vec{e}_i} = a^{y_d} \hat{\xi}_{t_{\sd} , \vec{x}_{\sd}}^{\epsilon_i \vec{e}_i} \ssp ,
\eea 
with $y_d \geq 0$ ($\xi_{t_{\sd} , \vec{x}_{\sd}}^{\epsilon_i \vec{e}_i}$ must be bounded) an exponent left undetermined for now, and the $\hat{\xi}_{t_{\sd} , \vec{x}_{\sd}}^{\epsilon_i \vec{e}_i}$ O(1) random variables satisfying the conservation of probability. The potential continuous noise is thus
\bea
\xi(\vec x , t) =\frac{D a^{y_d}}{A a^2}  \tilde{\xi}_{t_{\sd}=\frac{D}{a^2} t,\vec{x}_{\sd}=\Lambda^{-1} \frac{\vec x}{a}} \quad , \quad   \tilde{\xi}_{t_{\sd}, \vec x_{\sd}}  =\sum_{i=1}^d \sum_{\epsilon_i = \pm 1}  (B_i^{\epsilon_i} -1) \hat{\xi}_{\frac{a^2}{D_d} t,a \Lambda \vec x}^{\epsilon_i \vec e_i}   \ssp , 
\eea
as can be read off from (\ref{SM:EQ:Derivation2}). To understand its behavior in the $a \to 0$ limit, we now compute the following generating function, defined for an arbitrary test-function $f(\vec x , t)$ 
\bea \label{SM:GeneratingFunction}
\overline{e^{\int_{0}^{1} dt \int_{[0,1]^d} d^d x \xi(\vec x , t) f(\vec x , t)}}&&  = \overline{  e^{\frac{2^d a^{2+d} {\rm det}(\Lambda )}{ D  } \sum_{t_{\sd}=0}^{\frac{D}{a^2}} \sum_{x_1/2= 0 }^{\frac{\Lambda_1^{-1}}{2a}} \cdots \sum_{x_{d}/2= 0 }^{\frac{\Lambda_d^{-1}}{2a}} \frac{D a^{y_d}}{A a^2}   \tilde{\xi}_{t_{\sd},\vec x_{\sd}} f(2 a \Lambda \vec x_{\sd} , \frac{a^2}{D} t_{\sd} ) } } \nn \\
&&  = e^{\sum_{p=2}^{\infty}  \frac{1}{p!}\left(\frac{2^d a^{d+y_d} {\rm det}(\Lambda )}{ A} \right)^p  \sum_{t_{\sd}=0}^{\frac{D}{a^2}} \sum_{x_1/2= 0 }^{\frac{\Lambda_1^{-1}}{2a}} \cdots \sum_{x_{d}/2= 0 }^{\frac{\Lambda_d^{-1}}{2a}} \overline{ \tilde{\xi}^p}^c (f(2 a \Lambda \vec x_{\sd}, \frac{a^2}{D} t_{\sd} ))^p } \nn \\
&&  =  e^{\sum_{p=2}^{\infty}  \frac{1}{p!}\left(\frac{2^d a^{d+y_d} {\rm det}(\Lambda )}{ A} \right)^p \overline{ \tilde{\xi}^p}^c \frac{D }{2^d a^{2+d}  {\rm det}(\Lambda )}  \int_{0}^{1} dt \int_{[0,1]^d} d^d x (f(\vec x , t))^p}  \nn \\
&& = e^{\frac{1}{2} \sigma_{{\rm DP}}^2 \int_{0}^{1} dt \int_{[0,1]^d} d^d x  (f(\vec x , t))^2} + \cdots \ssp,
\eea
where in the last line we have retained the dominant order. $\xi(\vec x , t)$ is thus interpreted as a Gaussian white noise of strength $\sigma_{{\rm DP}}$ with
\bea \label{SM:SigmaDPDiscrete}
&& \overline{ \xi(\vec x , t) \xi(\vec x',t')  } = \sigma_{{\rm DP}}^2 \delta(t-t') \delta^d(\vec x- \vec x')  \nn \\
&& \sigma_{{\rm DP}}^2= a^{d-2 + 2 y_d }\frac{ 2^{d} D {\rm det}(\Lambda )}{A^2 } \overline{\tilde{\xi}^2} =  a^{d-2 + 2 y_d } \frac{ 2^{d} D {\rm det}(\Lambda )}{A^2 }  \overline{\left(\sum_{i=1}^d \sum_{\epsilon_i = \pm 1}  (B_i^{\epsilon_i} -1) \hat{\xi}_{\frac{a^2}{D_d} t,a \Lambda \vec x}^{\epsilon_i \vec e_i}  \right)^2}
\eea
as is presented in the letter. In dimension $1$ this result gives, noting $f_{\sd} = f_{\sd}^{\vec e_1} $
\bea \label{SM:ResSigmaDPDim1}
\sigma_{{\rm DP}}^2 = \frac{32 \sqrt{2} D f_{\sd}^2}{(1-4f_{\sd}^2)^2} a^{-1 + 2 y_1} \overline{\hat{\xi}^2}\ssp .
\eea
Hence, taking $y_1 = 1/2$ (weak noise scaling) we obtain a finite limiting noise and in this case the convergence of the discrete model to the MSHE should be rigorous. This result is similar but different from \cite{CorwinRWREWeakU} where the setting is for continuous limit of {\it non biased} random walk in a diffusive scaling but {\it not} around the diagonal of the lattice. A non trivial check of this result is given in the next section.

\medskip

Let us now also check how the general formula (\ref{SM:SigmaDPDiscrete}) compares with the result presented in the letter in the continuous setting, (\ref{Eq:DPNoiseStrength}). As explained in the previous section, the discrete setting formally reproduces the continuous one when the discrete model is rescaled diffusively in a low bias regime $f_{\sd} \sim a$. More precisely we take as in (\ref{SM:EQ:Rescalingfd}) $f_{d}^{ \epsilon_i \vec{e}_i} = a  \frac{1}{2D \sqrt{2d }}\epsilon_i \vec f \cdot \vec{e}_i $, $\xi_{t_{\sd}+1,\vec x_{\sd}}^{\vec e_i} = - \xi_{t_{\sd}+1,\vec x_{\sd}}^{-\vec e_i}$ and $\overline{\xi_{t_{\sd}+1,\vec x_{\sd}}^{\vec e_i}\xi_{t_{\sd}+1,\vec x_{\sd}}^{\vec e_j}} = \frac{\delta_{ij}}{ (2\sqrt{2d } )^{d+2}}$ (and thus $y_d = 0$). In that case we obtain $B_i^{\epsilon_i} \simeq 1 - \epsilon_i  \frac{\sqrt{2 d}}{D} a f_i$, $A \simeq 2$, $\tilde{B}_i \simeq 1$ and $\Lambda_i \simeq \sqrt{2d}$. We thus obtain from (\ref{SM:SigmaDPDiscrete})
\bea 
\sigma_{{\rm DP}}^2 && = a^{d-2 }\frac{ 2^{d} D (\sqrt{2d})^d}{4}  \sum_{i=1}^d (\frac{2\sqrt{2d}}{D} a)^2 f_i^2 \frac{1}{(2\sqrt{2d})^{d+2}}   \nn \\
&&  = \frac{f^2 a^d}{4 D} \ssp .
\eea
This indeed reproduces the continuum result (\ref{Eq:DPNoiseStrength}) if one identifies the lattice spacing $a$ with the characteristic length $r_0$ of the noise in the continuum setting.

\bigskip

{\bf Check of the weak-universality result using the Beta polymer}

Here we give a non-trivial check of the result (\ref{SM:ResSigmaDPDim1}) when the discrete noise is weak with $y_1 = 1/2$. To that aim we consider the Beta polymer, introduced in \cite{BarraquandCorwinBeta} and studied in \cite{BarraquandCorwinBeta,usBeta}. In the notation of the letter, the Beta polymer is obtained when taking the transition probability $p_{t_{\sd},x_{\sd}}\equiv p_{t_{\sd},x_{\sd}}^{\vec{e}_1} $ of the model of discrete random walk on $\JZ$ distributed as Beta random variables with parameters $(\alpha, \beta) \in \JR_+^2$. The moments of  $p_{t_{\sd},\vec{x}_{\sd}} \sim p$ are
\bea
\overline{p^n} = \frac{(\alpha)_n}{(\alpha+\beta)_n}
\eea
Hence, for $\beta = r \alpha $ with $r \neq 1$, the random walk is biased with
\bea
\overline{p}= 1/2+f_{\sd}= \frac{1}{1+r} \quad , \quad r = \frac{1/2-f_\sd}{1/2+f_\sd} \ssp .
\eea
In the large $\alpha$ limit, the fluctuations of the centered noise variable become small and Gaussian distributed:
\bea
p \sim \frac{1}{\sqrt{\alpha}} \hat p \ssp , 
\eea
with, for $\alpha \gg 1$, $\hat p$ distributed as a centered Gaussian random variable with $\overline{\hat p ^2} = \frac{r}{(1+r)^3}$. Comparing with the previous result in $d=1$, this means that the large $\alpha$ limit of the Beta polymer is formally equivalent to the diffusive weak noise limit of one-dimensional TD-RWRE on $\JZ$ with a lattice spacing $a$ given by $a = 1/\alpha$. More precisely we predict, following the previous section and the rescalings \eqref{Eq:SM:DiscreteRescaling11}-\eqref{Eq:SM:DiscreteRescaling12}, that in the Beta polymer, in the biased case $r \neq 1$, the limit
\bea \label{SM:ScalingLimitBeta}
Z_t(x) := \lim_{\alpha \to \infty} (2 \sqrt{(1/2+f_\sd)(1/2-f_\sd)})^{t_{\sd}} \left( \sqrt{ \frac{1/2-f_\sd}{1/2+f_\sd}} \right)^{x_{\sd}} P_{t_{\sd}}(x_{\sd})
\eea
with, on the right hand side $x_{\sd} = \frac{\alpha x}{\sqrt{2}} $ and $t_{\sd} = \alpha^2 t$, satisfies the MSHE
\bea
\partial_t Z_t(x) = \partial_x^2 Z_t(x)  + \xi_{{\rm DP}}(x,t)
\eea
with a Gaussian white noise of strength given by, using (\ref{SM:ResSigmaDPDim1})
\bea \label{SM:sigmaDPBeta}
\sigma_{{\rm DP}}^2 = \frac{(r-1)^2}{\sqrt{2} r (r+1)} \ssp .
\eea
Known results on the MSHE can thus be translated to results on the large $\alpha$ limit of the Beta polymer. In particular, using the result of \cite{droplet}, we predict the following Tracy-Widom asymptotics for the fluctuations of $\ln Z$ at large $t$:
\bea \label{SM:TracyWidomKPZ}
\lim_{t \to \infty} Proba \left( \frac{\ln Z_t(\phi t) +  (\frac{\sigma_{{\rm DP}}^4}{48} + \frac{\phi^2}{4} )t }{(\frac{\sigma_{{\rm DP}}^4}{4} t)^{1/3}}  <z \right) = F_2(z)
\eea
where $F_{2}$ is the cumulative distribution function of the Tracy-Widom GUE distribution.

\medskip

On the other hand in \cite{usBeta} we obtained an exact result similar to (\ref{SM:TracyWidomKPZ}) for the Beta polymer for arbitrary $\alpha,\beta$. Translated in the notations of this letter, the result of \cite{usBeta} reads, for $\frac{\beta-\alpha}{2(\alpha+\beta)} < \varphi <1/2$,

\begin{equation}\label{SM:TracyWidomBeta1}
\lim_{t_{\sd} \to \infty} Proba\left( \frac{ \ln Q_{t_{\sd}}(x_{\sd}=-2 \varphi t_{\sd}) + t_{\sd} c_{\varphi}}{(\lambda_{\varphi} t_{\sd})^{1/3} } < z \right) = F_2(z) \ ,
\end{equation}
with the parameters $c_{\varphi}$ and $\lambda_{\varphi} \sim t^{\frac{1}{3}}$ given by
\bea \label{SM:TracyWidomBeta2}
&& \varphi = \frac{\psi'(\beta + k_{\varphi})- \frac{1}{2} \left(\psi'(k_{\varphi}) + \psi'(\alpha+ \beta+ k_{\varphi}) \right)}{\psi'(\alpha+ \beta+ k_{\varphi}) -\psi'(k_{\varphi}) } \nn \\
&&  c_{\varphi} = \left(\varphi +\frac{1}{2}\right) \psi(k_{\varphi}+\alpha +\beta )-\psi(k_{\varphi}+\beta )+\left(\frac{1}{2}-\varphi \right) \psi(k_{\varphi})  \nn \\
 &&  2 \lambda_{\varphi}= -\left(\varphi +\frac{1}{2}\right) \psi''(k_{\varphi}+\alpha +\beta )+\psi''(k_{\varphi}+\beta )-\left(\frac{1}{2}-\varphi \right) \psi''(k_{\varphi}) \ .
\eea
Note in particular that the Beta polymer partition sum of \cite{usBeta} must be identified to the RW transition probability considered here as $Z_{t_\sd}(x_{\sd}) = Q_{t_{\sd}}(t_{\sd}-2 x)$. We have also changed, in the result (\ref{SM:TracyWidomBeta1})-(\ref{SM:TracyWidomBeta2}), $2^{2/3} \lambda_{\varphi} \to (\lambda_{\varphi} t_{\sd})^{1/3} $ to simplify the comparison with (\ref{SM:TracyWidomKPZ}).

\medskip
{}
Let us now compare (\ref{SM:TracyWidomBeta1})-(\ref{SM:TracyWidomBeta2}) and (\ref{SM:TracyWidomKPZ}) using the scaling limit (\ref{SM:ScalingLimitBeta}). Using the weak noise limit of the Beta polymer, the result (\ref{SM:TracyWidomBeta1})-(\ref{SM:TracyWidomBeta2}) should imply (\ref{SM:TracyWidomKPZ}). Let us thus take $r<1$ (the case $r>1$ can be treated similarly), $\beta = r \alpha$ with $\alpha \to \infty$ and scale $-2 \varphi =x_{\sd}/t_{\sd} = \frac{\phi}{\sqrt{2} \alpha}  $ in (\ref{SM:TracyWidomBeta1})-(\ref{SM:TracyWidomBeta2}). In this limit the system of equation (\ref{SM:TracyWidomBeta2}) is solved as
\bea
&& k_{\varphi}= \frac{\alpha  \left(-r^2-r\right)}{r-1}+\frac{r^2 \left(2 \sqrt{2} \phi +1\right)+2 r \left(\sqrt{2} \phi -1\right)+1}{2 (r-1)^2} + O(1/\alpha) \nn \\
&&  c_{\varphi}=  \frac{1}{2} \ln \left(\frac{1}{4 r}\right)+\ln (r+1) + \frac{\phi  \ln (r)}{2 \sqrt{2} \alpha }+\frac{\phi ^2}{4 \alpha ^2}+\frac{(r-1)^4}{96 \left(\alpha ^2 r^2 (r+1)^2\right)}  +  O(\frac{1}{\alpha^3}) \nn \\
&& \lambda_{\varphi} =\frac{(r-1)^4}{8 \alpha ^2 r^2 (r+1)^2} + O(\frac{1}{\alpha^3}) 
\eea
Thus in this limit the result (\ref{SM:TracyWidomBeta1})-(\ref{SM:TracyWidomBeta2}) predicts that the probability decays exponentially as
\bea
t_{\sd} c_\varphi = \alpha^2  t c_{\varphi} = t \left( \alpha^2\left( \frac{1}{2} \ln \left(\frac{1}{4 r}\right)+\ln (r+1) \right) + \alpha \frac{\phi  \ln (r)}{2 \sqrt{2} }  +\frac{\phi ^2}{4 \alpha ^2}+\frac{(r-1)^4}{96 \left(\alpha ^2 r^2 (r+1)^2\right) } \right) \ssp .
\eea 
This is indeed in agreement with (\ref{SM:ScalingLimitBeta}) and (\ref{SM:TracyWidomKPZ}). Indeed the term of order $O(\alpha)^2$ corresponds to the rescaling involving $ (2 \sqrt{(1/2+f_\sd)(1/2-f_\sd)})^{t_{\sd}}$ in (\ref{SM:ScalingLimitBeta}), the term of order $O(\alpha)$ corresponds to the rescaling involving $\left( \sqrt{ \frac{1/2-f_\sd}{1/2+f_\sd}} \right)^{x_{\sd}} $ in (\ref{SM:ScalingLimitBeta}), while the other terms are those predicted by (\ref{SM:TracyWidomKPZ}) and using (\ref{SM:sigmaDPBeta}). Finally the fluctuations of the logarithm of the probability are predicted to be given by
\bea
(t_{\sd} \lambda_{\varphi})^{1/3} \chi = (t \frac{(r-1)^4}{8  r^2 (r+1)^2})^{\frac{1}{3}} \chi \ssp ,
\eea
where $\chi$ is a random variable distributed according to the Tracy-Widom GUE distribution. This is again in agreement with (\ref{SM:TracyWidomKPZ}) using (\ref{SM:sigmaDPBeta}). We thus obtain a non-trivial check of our general weak-universality result in dimension $d=1$.

\bigskip
\newpage

{\bf On the forward equation} 

Here we consider the continuum TD-RWRE model of the Letter and detail the correspondence to the
KPZ equation in the forward setting, considering also here the possibility of applications to random environment with more general space time correlations. With the notations of the Letter, we consider the forward Kolmogorov (i.e. Fokker-Planck) equation for the probability $P(\vec x , t)$
that the particle is at position $\vec x$ at time $t$
\bea \label{EqFokkerPlanck}
&& \partial_t P(\vec x , t) =  D \nabla_x^2 P(\vec x , t) - \vec{\nabla}_x \cdot (( \vec f + \vec{\xi}(\vec x , t)) P(\vec x , t) )  \ssp 
\eea
where $\vec{\xi}$ is for the time being an unspecified noise term.
Here, for later applications, we are not specifying the initial condition, but if we are interested in the random-walk transition probability 
$P(\vec x , t) = P(\vec x , t | 0 ,0)$, we must choose the initial condition $P(t=0, \vec x) = \delta^d(x)$.

We now define $Z$ associated to the forward equation as
\bea  \label{Eq:SM:DefZForward}
Z(\vec x , t) := e^{  t \frac{ \vec f^2}{4 D} - \frac{\vec f \cdot \vec x}{2 D}}  P(\vec x , t) \ssp .
\eea
As in the backward case this rescaling is motivated by the fact that, if the noise is $\delta$-correlated in time, with Ito prescription,
the mean value of $P(\vec x , t)$ is the same as in the absence of noise, i.e. in the case of the random-walk transition probability (defined above),
$\overline{P(\vec x,t)} =  \frac{1}{(4 \pi D t)^{d/2}} e^{ - \frac{(\vec x-t \vec f)^2}{4 D t}}$. From \eqref{EqFokkerPlanck}, $Z$ satisfies the following stochastic equation
\bea
\partial_t Z(\vec x , t) =  D \nabla_x^2 Z(\vec x , t)  + \tilde{\xi}_{{\rm DP}}(\vec x , t) Z(\vec x , t) 
-  \vec{\xi}(\vec x , t) \cdot \vec{\nabla}_x  Z (\vec x , t) \ssp , \label{Eq:SM:blabla}
\eea  
where we have introduced a new DP noise term
\bea \label{newnoise} 
\tilde{\xi}_{{\rm DP}}(\vec x , t) := -  \frac{\vec f \cdot \vec{\xi}(\vec x , t)}{2D}  - \vec{\nabla}_x \cdot \vec{\xi}(\vec x , t) \ssp .
\eea
We thus almost obtain the same equation as in the backward case \eqref{Eq:SHE1}, up to an immaterial change
of sign in the last term, except that now
the forward `DP-noise' , as compared to the backward `DP-noise' $\xi_{DP} = - \frac{\vec f \cdot \vec{\xi}}{2D}$,
contains one more term. Note that the total difference with the MSHE arising from the backward equation
is a term $\vec{\nabla}_x \cdot (\vec{\xi}(\vec x , t) Z (\vec x , t) )$, for which the argument of irrelevance at large scale 
presented in the Letter should apply equally well. This is necessary for consistency since for statistically space-time translationally invariant random environments, as is considered here, one has the equality in law between forward and backward transition probability at the level of one-point (fixed $t$ and $\vec{x}$) observables: $P(t,\vec{x}) \sim Q(t, -\vec{x})$. Whatever is true in the backward case is thus also true in the forward one, at least for what concerns one-point statistics.

Until now Eqs. \eqref{Eq:SM:blabla}-\eqref{newnoise} formally hold for arbitrary noise $\vec \xi$. 
Let us now further specify the model. Consider $\vec \xi$ to be centered Gaussian and $\delta$ correlated in
time 
\bea
\overline{\xi_i(\vec x,t)  \xi_j(\vec x',t')} = D R_1(\frac{\vec x-\vec x'}{r_c}) \delta(t-t') \delta_{ij}
\eea 
which can be seen as the model in the text with $\tau_c=0$, $\sigma^2_\xi \int ds R_2(s)=1$ and with the Ito prescription.
Note that adding short-ranged correlations in time does not change the conclusions,
up to minor changes in length scales. Here $R_1$ is an unspecified dimensionless function. 

Let us now compare the DP noises appearing in the backward ($\xi_{{\rm DP}}$ given in \eqref{Eq:DPNoiseDef}) and forward cases $\tilde{\xi}_{{\rm DP}}$, which are centered Gaussian and correlated as 
\be
\overline{\xi_{{\rm DP}}(\vec{x} , t) \xi_{{\rm DP}}(\vec{x}' , t')} = \delta(t-t') R_{{\rm DP}}\left( \frac{\vec x- \vec x'}{r_c}\right) \quad , \quad \overline{\tilde{\xi}_{{\rm DP}}(\vec{x} , t) \tilde{\xi}_{{\rm DP}}(\vec{x}' , t')} = \delta(t-t') \tilde{R}_{{\rm DP}}\left( \frac{\vec x- \vec x'}{r_c}\right)
\ee 
with respectively the two functions $R_{{\rm DP}}$, $\tilde R_{{\rm DP}}$ (with dimensions $[\text{time}]^{-1}$)
\bea
&& R_{{\rm DP}}( \vec y) = \frac{f^2}{4D} R_1( \vec y) \ssp , \nn \\
&& \tilde{R}_{{\rm DP}}( \vec y) = \frac{f^2}{4D} R_1(\vec y)  + \frac{1}{2 r_c} \vec{f} \cdot \vec{\nabla}_y R_1(\vec y) + \frac{D}{r_c^2}\nabla_y^2 R_1(\vec y) \ssp .
\eea
functions of the dimensionless argument $\vec y$. 

Now, if we consider TD-RWRE with disorder LR correlated in space, i.e. such as power law correlated
at large distance, $R_1(\vec y) \sim |y|^{\alpha}$, the two noises (backward and forward) will have the same 
correlator at large distance and lead to the same mapping to the LR correlated noise KPZ equation mentionned in the text.
When the LR nature of these correlations become relevant for the KPZ equation itself, one will thus observe
the LR-KPZ universality classes. Indeed, similar arguments as for the SR-KPZ case will lead to irrelevance 
of the common additional term $-  \vec{\xi} \cdot \vec{\nabla}  Z$ to both equations. We will not explore
further this case here. These spatial LR-KPZ classes have been reexamined recently 
in \cite{KardarLR}.

In the case of SR disorder, i.e. $R_1$ a quicly decaying function on a scale of order unity, we observe that the two correlators
differ only by derivatives, hence
\bea
\int d^d \vec{y} \, R_{{\rm DP}}( \vec y) =\int d^d \vec{y} \, \tilde{R}_{{\rm DP}}( \vec y) 
\eea 
It is well known in the DP context \cite{BustingorryDoussalRosso} that in the weak
noise limit in $d=1,2$, the KPZ equation in presence of a SR correlated noise becomes equivalent, beyond some scale, to a 
model with delta correlations in space and that the only information that is retained about the noise at large scale 
is the integral over space of the noise correlations. That shows once again that the backward and
forward equations lead to equivalent descriptions. 

Let us be more precise in $d=1$. Consider the forward 
equation \label{Eq:SM:SHEForward2} in units $x^*$ and $t^*=(x^*)^2/D$ (see next section for details). 
In these units the correlator of the "DP noise" becomes
\bea \label{coarse} 
R_{{\rm DP}}\left( \frac{x'}{r_c}\right)  \to  \frac{(x^*)^2}{D} \frac{f^2 }{4 D} R_1\left(x^* \frac{x'}{r_c}\right) \to_{x^*/r_c \gg 1} 2 \delta(x') 
\eea 
where we have used $x^* = \frac{8 D^2}{r_0 f^2}$ and $r_0=r_c \int dy R_1(y)$ (as defined in the text).
Clearly the replacement by a $\delta$ correlated noise is valid if the scale of the EW-KPZ crossover
(see next section) $x^* = \frac{8 D^2}{r_0 f^2} \gg r_c$,
where $r_c$ is the range of the disorder. A similar estimate as \eqref{coarse}
shows that $R_{{\rm DP}}( \frac{x}{r_c}) \simeq \tilde R_{{\rm DP}}( \frac{x}{r_c})$ at 
scales $x \gg r_c$, i.e. at this scale the two DP noises (backward and forward) are equivalent. 
The difference being a total derivative, it spatially averages quickly to zero when coarse-grained at scales
larger than $r_c$.

This leaves the question of the scale above which one can neglect the non-KPZ term $-  \vec{\xi} \cdot \vec{\nabla}  Z$.
As mentionned in the text a simple estimate for this scale is $x_f = D/f$. Note that in the forward setting
it is also possible to consider cases where the initial probability $P(x, t=0)$ is spread over a region $\delta x=w$.
If $w>x_f$ the effect of the non-KPZ term is negligible at all scales.

\bigskip

{\bf Exact result in the `small large deviations' regime in the continuum in $d=1$}

We consider the continuum TD-RWRE model of the letter and $h(x,t) = \ln Z(x,t)$ with $Z(x,t)$ given by the rescaled backward probability as in \eqref{Eq:DefZ}. The `height' $h(x,t)$ latter satisfies \eqref{Eq:KPZ1} and, as mentioned in the letter, rescaling time, space and height in dimension $1$ as $t = t^* t'$, $x = x^* x'$ and $h'(t',x') := \frac{1}{h^*} h(t^* t' , x^* x')$ with the characteristic scales $t^* = \frac{(4 D)^3}{r_0^2 f^4}$,
$x^* = \frac{8 D^2}{r_0 f^2} $ and $h^*=1$ leads to an equation for $h'$ that reads (the same rescaling for the KPZ equation \eqref{KPZ1} to bring it to $D_0=\lambda_0=2$, $\nu=1$ lead the characteristic scales $x^*=(2 \nu_0)^3/(D_0\lambda_0^2)$, 
$t^*=2(2\nu_0)^5/(D_0^2\lambda_0^4)$ and $h^* = (2\nu_0)/\lambda_0$).
\be
\partial_{t'} h'(x',t) =  \partial_{x'}^2 h'(x',t') + (\partial_{x'} h'(x',t'))^2 + \sqrt{2} \eta(x',t')  + \frac{r_0 f}{2 \sqrt{2} D} \eta(x',t') 
\partial_{x'} h'(x',t')  \label{KPZ resc}
\ee
with $\eta$ a unit Gaussian white noise, 
$\langle \eta(x_1',t'_1) \eta(x_2',t'_2) \rangle = \delta(t'_1-t'_2) \delta(x'_1- x'_2)$. Note that this holds equally well in the forward case with $Z(x,t)$ given by the rescaled forward probability (\ref{Eq:SM:DefZForward}) and $h(x,t) = \ln Z(x,t)$, the additional noise
term then becomes $\frac{r_0 f}{2 \sqrt{2} D} \partial_{x'}( \eta(x',t')  h'(x',t'))$.
In both cases, this means that, up to sub-dominant corrections $h'(x',t')$ 
solves the KPZ equation. At large $t'$ we may use the result of \cite{droplet} which reads
\bea \label{Eq:SM:ResDroplet}
\lim_{t' \to \infty} \frac{h'(x',t') + \frac{t'}{12}+  \frac{x'^2}{4 t'} }{(t')^{1/3}} = \chi_2 \ssp ,
\eea
with $\chi_2$ a RV with a TW GUE distribution. 

This result is valid in the regime of a weak bias $f$. The weak bias is naturally obtained by looking at `small large deviation' around the optimal direction. In the presence of a systematic bias $f$, the optimal direction in the forward setting is for $x = ft + o(t)$. Looking around the direction $x  = (f + u)t + x'$ introduces an effective bias $f_u = - u$ (here the change of sign compared to the effective bias introduced in the backward setting in the letter is due to the fact that time is counted negatively in the backward setting) and we should thus write
\bea
P((f+u)t + x , t) =  e^{- \frac{u^2 t}{4D} - \frac{x u }{2D} + h(x,t)} \ssp .
\eea
Introducing $h'(t',x') := \frac{1}{h^*} h(t^* t' , x^* x')$ with $h^* = 1$, $x^*=  \frac{8 D^2}{r_0 u^2} $ and  
$t^* = \frac{(4 D)^3}{r_0^2 u^4}$ we have that for small $u$ and large $t'$, $h'$ is distributed as \eqref{Eq:SM:ResDroplet}. Combining everything we conclude that
\bea \label{Eq:SM:ResDropletLogP}
\lim_{t \to \infty}  \frac{\ln P((f+u)t + x , t) + I_{\rm q}(u) t}{\lambda(u) t^{1/3} } = \chi_2 \ssp , 
\eea
with
\bea \label{Eq:SM:ResIqAndLambda}
&& I_{\rm q}(u) = \frac{u^2}{4D} + \frac{h^*}{12 t^*} =  \frac{u^2}{4D} + \frac{2 r_0^2 u^4}{3 (8 D)^3} \nn \\
&& \lambda(u) = \frac{h^*}{(t^*)^{1/3}} = \frac{r_0^{2/3} u^{4/3}}{4D}  \ssp ,
\eea
which is the result presented in the letter in the absence of a bias \eqref{Eq:ResDropletLetter}. This result is valid at small $u$: \eqref{Eq:SM:ResDropletLogP} should be valid in full generality and \eqref{Eq:SM:ResIqAndLambda} gives the first terms in the expansion of the quenched large deviations function $I_{\rm q}(u)$ and of the amplitude of the fluctuations $\lambda(u)$, and corrections arise when $u \simeq D/r_0$, beyond which the small scale dependence of the
noise starts to matter. 

Let us now explore further the consequences of the mapping of the backward diffusion equation (respectively forward)
on \eqref{KPZ resc} in the intermediate time regime. We know that if the last term in \eqref{KPZ resc} can be neglected then
at a given $(x'=x/x^*,t'=t/t^*, h^*=1)$ we have the equality in law
\bea 
&& h(x,t)= h^* h'(x',t') = - \frac{t'}{12} + \chi^{KPZ}_{t'} \label{EWCross}  \\
&& \chi^{KPZ}_{t'} \simeq (t')^{1/3}  \chi_2  \quad , \quad t'=t/t^* \gg 1 \nn \\
&& \chi^{KPZ}_{t'} \simeq c_0 (t')^{1/4}  \omega  \quad , \quad t'=t/t^* \ll 1 \nn
\eea 
Here $\omega$ is a unit Gaussian variable, and the corresponding regime of small rescaled time is called the 
Edwards Wilkinson (EW) regime.
Here, for a droplet (i.e. localized in space) initial conditions $c_0=1/\sqrt{2 \pi}$ and
$\chi_2$ is a GUE TW random variable. In that case the exact PDF of the $O(1)$ random variable
$\chi^{KPZ}_{t'}$ is known \cite{droplet} for arbitrary $t'$, and crosses over
from Gaussian to GUE TW as $t'$ increases. For different
initial conditions, e.g. flat, $\chi_2$ is replaced by another universal distribution, and $c_0$ takes
a different value. However the general form of the above result is universal for $t \geq t_c$ where $t_c$ is 
some small cutoff time. In general, for the KPZ equation, $t_c$ depends on details of the noise and of
the initial condition, but here it also depends on the neglected term.
Hence, to be on the safe side, we assume that \eqref{EWCross} 
is valid for $t \gg t_f=D/f^2$, where $t_f$ is the scale identified above
below which the non-KPZ term becomes relevant. Thus for
the EW regime to exist, and the EW-KPZ crossover to be observable, we need $t^* \gg t_f$, 
i.e. $r_0 f/(8 D) \ll 1$,
as stated in the text,
i.e. to be in the small bias regime. Let us note that at $t=t_f$ the magnitude of the
(Gaussian) fluctuations from \eqref{EWCross} are $\sim \sqrt{r_0 f/(8 D)}$. 

Some of the conclusions which can be extracted
from \eqref{EWCross} for the diffusion problem have been discussed in the text,
such as the existence of an EW intermediate time regime. 
Let us come back to the forward setting discussed above i.e. 
\bea \label{PEW} 
P((f+u)t + x , t) = - \frac{u^2 t}{4D} - \frac{x u }{2D} + h(x,t) =  - \frac{u^2 t}{4D} - \frac{x u }{2D} 
- \frac{t}{12 t^*} + \chi^{KPZ}_{t/t^*} 
\eea
where we must identify, as discussed above, $f \to f_u = f-(f+u)$ hence $x^*=  \frac{8 D^2}{r_0 u^2} $ and  
$t^* = \frac{(4 D)^3}{r_0^2 u^4}$ (the case $u=-f$ corresponds to the first case studied in the Letter).
Eq. \eqref{PEW} is correct, in law, for fixed $x$, and for fixed $u$ and $t/t^*$ when $t^* \gg D/u^2 $.
 Let us set $x=0$. Let us denote
$X= u t$ the distance from the ``optimal probability", i.e. the maximum of $P$. We can rewrite
\bea \label{PEW2} 
P(f t + X , t) =   - \frac{X^2}{4D t} - \frac{t}{12 t^*} + \chi^{KPZ}_{t/t^*}  , 
\eea
and the first term is safely interpreted as the bulk of the probability, i.e. normal
diffusion, and \eqref{PEW2} thus also reproduces (at least the deterministic part of) the $X=O(\sqrt{t})$ regime where $u=O(1/\sqrt{t})$. The other term represent the corrections due to the disorder
and we want to estimate the magnitude of its fluctuations. Let us denote
\bea \label{ydef} 
X = y \, r_0 (\frac{4 D t}{r_0^2})^{3/4} 
\eea 
where $y$ is a dimensionless parameter. Then we see that 
\bea
\frac{t}{t^*} = y^4  \ssp .
\eea 
Hence fixed $t/t^*$ means fixed $y$. We now assume that \eqref{PEW} 
is valid for any $u$ (at fixed $t/t^*$) and even for $u = O(t^{-1/4})$, in between the diffusive scaling $u = O(1/\sqrt{t})$ and the large deviation regime $u=O(1)$. Under that
assumption we thus find that \eqref{PEW2} is valid for fixed value of $y$ and we obtain using \eqref{PEW2} with $u \to X/t$
\bea \label{PEW33} 
P(f t + X , t) =   - \frac{X^2}{4D t} - \frac{y^4}{12}  + \chi^{KPZ}_{y^4} 
\eea
Using \eqref{EWCross} we see that the crossover from EW to KPZ occurs on scales such that $X \sim t^{3/4}$
as announced, as $y=O(1)$ increases. In the EW regime $y \ll 1$ we have
\bea \label{PEW3} 
P(f t + X , t) \simeq_{y \ll 1} - \frac{X^2}{4D t}  + c_0 y \, \omega
\eea
The crossover to diffusion occurs when considering $X \sim \sqrt{4 D t}$. Then the fluctuations in 
\eqref{PEW3} are Gaussian and of magnitude
$y \sim (\frac{r_0^2}{4 D t})^{1/4}$. It also corresponds to the lower edge of
validity of \eqref{EWCross} since $X\ = O(x_f = \frac{D t}{X})$.
It is easy to see that such fluctuations can be interpreted
as some local averaging, within a diffusion volume $x \sim \sqrt{D t}$, of the disorder as
\be
\ln P \sim \frac{1}{D t} \int_0^t dt \int_{x \sim \sqrt{D t}} dx \, \xi(x,t) \sim (\frac{r_0}{\sqrt{D t}})^{1/2} \omega
\ee
Thus the full crossover from diffusion $X \sim t^{1/2}$, through intermediate $X \sim t^{3/4}$, to large
deviations $X \sim t$ can be accounted by the present analysis.

\bigskip

{\bf Details on the simulations}

The observation of GOE type statistics reported in the letter was made using simulations of a discrete model of RWRE on $\JZ \cap [-L/2,L/2]$ with reflexive boundary conditions and $L=8192$. The random environment was chosen such that the random hopping probabilities are given by $p_{x_\sd,t_\sd} = (y_{x_\sd,t_\sd})^{1/3}$ with the $y_{x_\sd,t_\sd}$  iid random variables. This introduces a bias with $f_{\sd}:= \overline{p_{x_\sd,t_\sd}}-1/2=1/4$. At each time step a particle initially located at $x_{\sd}$ jumps to the right with probability $p_{x_\sd,t_\sd}$ and to the left with probability $1-p_{x_\sd,t_\sd}$. Particle attempting to jump out of the system just remains at their position (reflexive boundary condition). We start at the initial time with a distribution of particle $P_{t_\sd = 0}(x_\sd)$ given by
\bea
P_{t_\sd=0}(x_\sd) = B^{x_\sd}/{\cal N}
\eea
with $B= \frac{1/2 + f_{\sd}}{1/2- f_{\sd}}$ and ${\cal N}$ a normalization constant. The probability is then evolved as
\bea
&& P_{t_\sd+1}(x_\sd) = p_{x_\sd-1,t_\sd} P_{t_\sd}(x_\sd-1)  +(1-p_{x_\sd+1,t_\sd})P_{t_\sd}(x_\sd+1) \quad \forall  x_{\sd} \in \{ -L/2 +1 \cdots L/2-1 \}  \nn \\
&& P_{t_\sd+1}(L/2) = p_{L/2-1,t_\sd} P_{t_\sd}(L/2-1) +p_{L/2,t_\sd} P_{t_\sd}(L/2)  \nn \\
&&  P_{t_\sd+1}(-L/2) = (1-p_{-L/2+1,t_\sd} )P_{t_\sd}(-L/2+1) +(1-p_{-L/2,t_\sd}) P_{t_\sd}(-L/2) \ssp .
\eea
The initial condition is chosen so that in the absence of disorder (putting $p_{x_\sd,t_\sd} \to 1/2 +f_\sd $ above) it is invariant: $\overline{P}_{t_\sd}(x_\sd) = P_{t_\sd=0}(x_\sd) $. This is a discrete analogue of the setting (presented in the continuum in the letter) were GOE type fluctuations for the logarithm of the transition probability are expected. Statistics were obtained using $4 \times 10^6$ simuations and for each one we store $\ln P_{t_\sd}(0)$ for $t_{\sd} \in \{128,256,512,1024,2048\}$. In Fig~\ref{fig:GOE} we show the comparison between the histogram of the (normalized and centered) values of $\ln P_{2048}(0)$ and compare them to the (centered and normalized) GUE and GOE TW distribution. The insets also show the convergence (with increasing $t_{\sd}$) of the skewness and kurtosis of the  distribution of $\ln P_{t_\sd}(0)$ to that of the GOE TW distribution, allowing an unambiguous distinction with the GUE values.

\bigskip

{\bf Position of the maximum of a large number of independent random walkers}

Here we give the derivation of formula (\ref{Eq:Xmax}). Similar arguments can be found in the proof of Corollary 5.8 of \cite{BarraquandCorwinBeta}.

\smallskip

Using the notations of the letter, we consider for a non-biased TD-RWRE, $\hat{P}(x,t) = 
{\rm Prob}(x_1(t) >x~ |\vec x(0) = 0)$, the probability that a random walker in dimension $d$ starting at time $0$ at position $0$ ends up up at time $t$ in the region $\Omega_x = \{ \vec{x} \in \JR^d, \vec{x} \cdot \vec{e}_1 \geq x \}$. Following the discussion in the letter we conjecture that for $u$ large enough ($u>u_c$)
 \bea \label{SM:Max:TW1}
\chi:= \lim_{t \to \infty} \frac{\ln \hat{P}(u t,t) + I_{{\rm q}}(u) t }{ \lambda(u) t^{\theta_d}} \ssp , 
\eea
defines a $O(1)$ random variable with a universal distribution related to the KPZUC. In the usual DP terminology
this would be a point to half-space distribution. However, since this is a geometry where the DP is stretched (large
deviation regime) we expect this probability to coincide with the point to hyperplane (of dimension $d-1$) KPZ 
subuniversality: $\ln {\rm Prob}(x_1(t) > u t~ |\vec x(0) = 0) \sim
\ln {\rm Prob}(x_1(t) \approx u t ~ |\vec x(0) = 0)$ (e.g. point to point in $d=1$, point to line in $d=2$, point to plane in $d=3$). Such distributions
have been studied in $d=2$ in \cite{halpin2012-2D}). Here $\lambda(u)$ and $I_{{\rm q}}(u)$ are some non-universal constants ($I_{{\rm q}}(u)$ is the quenched large deviations function of the RW) that permits to center and rescale $\ln \hat{P}(t, u t)$ according to the usual KPZ scaling. Here $u$ should be larger than some critical velocity, $u>u_c$, in dimension $d >2$, while $u>0$ is sufficient in $d=1,2$ ($u_c=0$).

We consider now $N \gg 1$ random walkers diffusing in the same time-dependent random environment and starting at the origin at time $0$. We are interested in the cumulative probability of the maximum distance travelled by the random walkers in the direction defined by the unit vector $\vec{e}_1$, $x_{{\rm max}} := {\rm max}_i \{\vec{x}_i(t) \cdot \vec{e}_1 \}$. Here by probability we now mean the probability to observe a certain outcome for one experiment in one environment. We can write
\bea
{\rm Prob}(x_{{\rm max}} \leq u t ) && = \overline{\prod_{i=1}^N ( 1- {\rm Prob}( \vec x(t) \cdot \vec{e}_1 \geq ut | \vec x(0) = 0 ))} \nn \\
&&  = \overline{(1 - \hat{P}(t, u t))^N } = \overline{e^{N \ln( 1 - \hat{P}(t, u t))} } = \overline{e^{ - N  \hat{P}(t, u t)} }\nn \\
&& = \overline{\exp\left( - e^{\ln(N) + \ln(\hat{P}(t, u t))}\right)} = \overline{\exp\left( - e^{\ln(N) - t I_{{\rm q}}(u) + \lambda(u)t^{\theta_d} \chi}\right) } \label{argumentN} 
\eea
Here we have taken the dominant behavior for $t$ large and $u>u_c$ so that we may use (\ref{SM:Max:TW1}). This is ensured if we choose as in the letter $t \gg 1$ and $N \gg 1$ with $\gamma = \ln(N)/t $ fixed, such that 
\be
u_\gamma^* := I_{{\rm q}}^{-1}(\gamma) \Leftrightarrow  I_{{\rm q}}(u_\gamma^*)= \gamma \quad , \quad \gamma = \frac{\ln N}{t} \quad \text{fixed} 
\ee
and $u_\gamma^* >u_c$. Indeed, in that case we obtain, taking
\bea
u = u_\gamma^* + c(\gamma) t^{\theta_d } \frac{\delta \tilde x}{t} \ssp  \quad , \quad c(\gamma) 
= \frac{\lambda(u_\gamma^*)}{I_{{\rm q}}'(u_\gamma^*)} >0
\eea
with $\delta \tilde x$ a rescaled displacement, 
\bea
{\rm Prob}(x_{{\rm max}} \leq u_\gamma^*t  + c(\gamma) t^{\theta_d } \delta \tilde x )  = \overline{ \exp\left( - e^{\lambda(u)t^{\theta_d} ( \chi - \delta \tilde x) + o(t^{\theta_d})} \right)}   \sim_{t \to \infty}  \overline{ \theta(\delta \tilde x-\chi) } \ssp .
\eea
Which shows that, in law
\bea \label{Eq:SM:xmaxDistribution}
x_{{\rm max}} \simeq u_\gamma^*t  + c(\gamma) t^{\theta_d } \chi \ssp .   \label{max1} 
\eea
Hence $x_{{\rm max}}$ is located around a ballistically moving front $u_\gamma^* t$ but has large, random environment-dependent fluctuations around it, scaling with the KPZUC exponent $t^{\theta_d}$ and whose distribution is universal. The precise expected distribution for $\chi$ has been discussed above. Here the argument is valid in any $d$ and for arbitrary models, including continuum or lattice RW models, provided $u^*_{\gamma} > u_c$ (with $u_c=0$ in $d=1,2$).

As emphasized above, this formula gives the dominant behavior of the distribution of $x_{{\rm max}}$ for one experiment in one random environment. If one repeats the experiment many times in the {\it same} time-dependent environment, the random variable $\chi$ in the above formula is now fixed and the fluctuations of $x_{{\rm max}}$ of the Gumbel type, as should be expected from the general theory of extreme value statistics. More precisely we can write 
\be
x_{{\rm max}} = \langle x_{{\rm max}} \rangle + \frac{G - \gamma_E}{I'_{{\rm q}}(u_\gamma^*)} \label{gum1} 
\ee
where $\langle x_{{\rm max}} \rangle$ is the thermal average of the position of the maximum.
The latter is a random variable which fluctuates from sample to sample as \eqref{max1} (up to
subdominant terms), and $G - \gamma_E$ is a centered standard Gumbel random variable, independent from 
$\chi$, which gives the leading behavior of the thermal fluctuations. Note that to
observe in practice these Gumbel fluctuations requires considering
$m \gg 1$ groups of $N$ independent particles and investigating the group to group
fluctuations.

Let us finally use the exact result  \eqref{Eq:SM:ResDropletLogP} in the weak bias limit in the continuum in $d=1$ to obtain a sharp prediction: in that case $\theta_d=1/3$, and the functions $I_{{\rm q}}(u)$ and $\lambda(u)$ were given in \eqref{Eq:SM:ResIqAndLambda}.
We find (to leading order)
\be
u_\gamma^* = \sqrt{4 D \gamma} \left( 1-  \frac{\gamma r_0^2}{96 D} + o((\frac{\gamma r_0^2}{D})^{2}) \right)  \quad , \quad c(\gamma) \simeq 
\frac{1}{2} r_0^{2/3} (u_\gamma^*)^{1/3} \simeq 
\frac{1}{2} r_0^{2/3} (4 D \gamma)^{1/6}   \ssp , 
\ee
to leading orders in an expansion in $r_0 u_\gamma^*/D \ll 1$. Note that $u_\gamma^* t \simeq \sqrt{4 D \gamma} = \sqrt{4 D t \ln N}$ which is indeed the leading
behavior (in $N$) of the maximum of $N$ independent Gaussian random variables
corresponding to the diffusion (a result also correct for fixed $t$ and large $N$).
Note also that inserting $u_\gamma^*$ in \eqref{gum1} recovers the standard
result $G\sqrt{4Dt}/\sqrt{\ln N}$ for the fluctuations of the maximum $N$ independent Gaussian random variables
corresponding to the diffusion.

A more refined analysis in the continuum in $d=1$ shows that the condition $\ln N \sim t$ can be relaxed a bit and
that the KPZ fluctuations are observed as long as
\be
\ln N \gg (\frac{4 D t}{r_0^2})^{1/2}  \label{condN2} 
\ee
i.e. $\ln N \sim t^{1/2}$ is the borderline case. Eq \eqref{condN2} implies
that the leading behavior $x^0_m$ of the position of the maximum satisfies
\be
x^0_m := \sqrt{4 D t \ln N} \gg r_0 (\frac{4 D t}{r_0^2})^{3/4} 
\ee 
Comparing with \eqref{ydef} shows that this is the criterion for
being in the KPZ regime, i.e. fluctuations of $\ln \hat P$ being large
in \eqref{argumentN}. This more general result thus reads
\bea \label{Eq:SM:xmaxDistribution2}
x_{{\rm max}} \simeq \sqrt{4 D t \ln N}  + \frac{1}{2} r_0^{2/3} (4 D t \ln N)^{1/6} \chi
\eea
where, for a localized initial condition, $\chi=\chi_2$ the GUE TW random variable. 
It recovers \eqref{gum1} for $\ln N=\gamma t$, but its domain of validity is broader 
and given by the condition \eqref{condN2}. Note that the other condition is
that $\ln N \gg 1$ so that $x^0_m \gg \sqrt{D t}$ (which then implies $t \gg t_u$).

\bigskip

{\bf Hitting time distribution}

A related question is to obtain the hitting time distribution far from the most probable direction of the diffusion.
Given the region $\Omega_\ell = \{ \vec x \in \JR^d, \vec x \cdot \vec e_1 \geq \ell \}$,
we now ask what is the first time $T_{{\rm Hit}}(\ell)$ at which one particle of the
cloud of $N$ particles (initially localized around $\vec x=0$) will enter $\Omega_\ell$. 
For this problem one must take 
\bea
\hat{\gamma} := \frac{\ln N}{\ell}
\eea
fixed with $N, \ell \to \infty$. It is then clear that to leading order $T_{{\rm Hit}}(\ell)$ is linear in $\ell$:
\bea \label{defImplicit} 
T_{{\rm Hit}}(\ell) \simeq \ell/v_{\hat \gamma}^*
\eea
with $v^*_{\hat{\gamma}}$ a characteristic velocity, which is the positive solution of the self-consistent equation,
\bea
I_{{\rm q}}(v_{\hat \gamma}^*) = \frac{\ln N}{ \ell}v_{\hat \gamma}^*  = \hat{\gamma} v_{\hat \gamma}^* \ssp .
\eea
This determines uniquely $v^*_{\hat \gamma}$ since we always have $I_{{\rm q}}'(0) =0$ and $I_{{\rm q}}(v) \geq I_{{\rm a}}(v) = \frac{v^2}{4D}$. One also expects fluctuations around this linear behavior and we write
\bea
T_{{\rm Hit}}(\ell) \simeq \ell/v_{\hat \gamma}^*  + \ell^{\alpha}  \delta \tilde{T} + o(\ell^{\alpha}) \ssp, 
\eea
with $\delta \tilde{T}$ some $O(1)$ fluctuations and $\alpha < 1$ an exponent determined below. Now we have by definition that $x_{{\rm max}}(T_{{\rm Hit}}(\ell)) = \ell \simeq x_{{\rm max}}(t= \ell/v^*  + \ell^{\alpha}  \delta \tilde{T})$. Using now that $x_{{\rm max}}(T_{{\rm Hit}}(\ell))$ is given by \eqref{Eq:SM:xmaxDistribution} with $u_\gamma^*$ and $\gamma$ such that
\bea
I_{{\rm q}}(u_\gamma^*) = \gamma =\frac{\ln N}{T_{{\rm Hit}}(\ell)} = \frac{\ln N}{\ell/v_{\hat \gamma}^*  + \ell^{\alpha}  \delta \tilde{T}} \simeq \hat{\gamma} v_{\hat \gamma}^* - \hat{\gamma}(v_{\hat \gamma}^*)^2  \ell^{\alpha-1} \delta \tilde{T}  \ssp , 
\eea
Note that the value of $\gamma$ one must use in \eqref{Eq:SM:xmaxDistribution} is not $\hat{\gamma} v_{\hat \gamma}^*$
but slightly shifted. Hence we find
\bea
u_\gamma^* \simeq v_{\hat \gamma}^*  - \frac{\hat{\gamma} (v_{\hat \gamma}^*)^2 \ell^{\alpha-1} \delta \tilde{T} }{I_{\rm q}'(v_{\hat \gamma}^*)} \ssp .
\eea
Inserting this expansion in $x_{{\rm max}}(T_{{\rm Hit}}(\ell)) =\ell \simeq u_\gamma^* ( \ell/v_{\hat \gamma}^*  + \ell^{\alpha}  \delta \tilde{T} ) + c(\gamma) ( \ell/v_{\hat \gamma}^*  +  \ell^{\alpha}  \delta \tilde{T} )^{\theta_d} \chi $ with $c(\gamma) = \frac{\lambda(u_\gamma^*)}{I_{\rm q}'(u_\gamma^*)}$ and expanding with $\alpha = \theta_d$ we obtain to leading order
at large $\ell$
\bea \label{reshitt} 
T_{{\rm Hit}}(\ell) \simeq \frac{\ell}{v_{\hat \gamma}^*} - d(\hat \gamma) \ell^{\theta_d} \chi + o(\ell^{\theta_d}) \quad , \quad 
d(\hat \gamma) = \frac{\lambda(v_{\hat \gamma}^*)}{(v_{\hat \gamma}^*)^{\theta_d+1}  ( I_{\rm q}'(v_{\hat \gamma}^*) - \hat{\gamma})} 
=  \frac{  \lambda(v_{\hat \gamma}^*)  \partial_{\hat \gamma}  v_{\hat \gamma}^*}{(v_{\hat \gamma}^*)^{\theta_d+2}} \ssp .
\eea
where in the last equality we have used that $I_{{\rm q}}'(v_{\hat \gamma}^*) - \hat{\gamma} = \frac{v_{\hat \gamma}^*}{ \partial_{\hat \gamma} v_{\hat \gamma}^* }$, obtained by differentiating the the implicit equation \eqref{defImplicit} defining $v_{\hat \gamma}^*$. 

 This heuristic calculation thus indicates that the hitting time distribution is also universal (exponent and distribution)
in the limit $\ell \to +\infty$ at fixed $\hat \gamma$.
Here the argument is valid in any $d$ and for arbitrary models, including continuum or lattice RW models, provided $\hat v^*_{\hat \gamma} > u_c$ (with $u_c=0$ in $d=1,2$). For the above choice of domain, $\Omega_\ell$, and initial condition, localized around $\vec x=0$, as discussed in the previous section,
$\chi$ is a ``point to co-dimension one" KPZ random variable: ``point to point", i.e. droplet GUE TW in $d=1$, ``point to line" in $d=2$, ``point to plane" in $d=3$ (see Fig.~\ref{fig:GeometriesPic}).

In the continuum $d=1$ model we can further estimate the prefactors. Using \eqref{Eq:SM:ResIqAndLambda} we obtain to leading order 
\bea
&& T_{{\rm Hit}}(\ell) \simeq \frac{\ell}{v_{\hat \gamma}^*} - d(\hat \gamma) \ell^{1/3} \chi_2 + o(\ell^{\theta_d}) \\
&& v_{\hat \gamma}^* = 4 D \hat \gamma \left(1 - \frac{\hat \gamma^2 r_0^2}{12} + O(\hat \gamma^3 r_0^3) \right) \\
&& d({\hat \gamma}) \simeq \frac{1}{4 D \hat \gamma} r_0^{2/3}  \quad , \quad \hat \gamma = \frac{\ln N}{\ell} 
\eea 
for $N,\ell \to +\infty$ at fixed $\hat \gamma$, to leading orders in an expansion in powers of $\hat \gamma r_0 \ll 1$. Here 
$\chi_2$ is a GUE TW random variable.

\bigskip

We now generalize this prediction for the hitting time to other domains. Consider in particular the interesting case of a bounded domain, and to fix the ideas consider $B(\ell) = \{ \vec{x} \in \JR^d , |\vec x - \ell \vec{e}_1| \leq  R\}$, a ball of radius $R$ centered on $\ell \vec{e}_1$ (we comment in the end of the section on the size of $R$ that should be considered for this calculation to be valid). Then one expects that for a single particle
\bea
\lim_{t \to \infty} \frac{\ln {\rm Prob}(\vec{x}(t) \in B(ut)|\vec x(0) = 0) +  t\hat{I}_{{\rm q}}(u)}{\hat{\lambda}(u) t^{\theta_d}} = \hat{\chi}
\eea
with $\hat{I}_{{\rm q}}(u)$ (=$I_{{\rm q}}(u)$ in that case) and $\hat{\lambda}(u)$ some velocity dependent constants and $\chi$ the universal distribution of the fluctuations of free-energy of DPs in dimension $d+1$ with both endpoints fixed. The probability that no particle among $N$ particles is present in $B(ut)$ at time $t$ is thus
\bea \label{Eq:SM:HitBall1}
\overline{(1-{\rm Prob}(\vec{x}(t) \in B(ut)|\vec x(0 = 0)))^N} \simeq \overline{e^{-N e^{- t \hat{I}_{{\rm q}}(u) +\hat{\lambda}(u) t^{\theta_d} \hat \chi}}}
\eea
Taking again $\hat \gamma = \frac{\ln N}{\ell}$ fixed with $\ell \to \infty$ we consider $T_{{\rm Hit}}(\ell)$ the first time that a particle enters into $B(\ell)$. We expect again the behavior 
\bea
T_{{\rm Hit}}(\ell) \simeq \ell/v^* + \ell^{\alpha} \delta \tilde{T}
\eea
with $v^*$ some characteristic velocity determined below and $\delta \tilde T = O(1)$ some fluctuations. A natural interpretation of \eqref{Eq:SM:HitBall1} in terms of hitting time probability is that, introducing $t = \ell/v^* + \ell^{\alpha} y $ in \eqref{Eq:SM:HitBall1} 
\bea
{\rm Prob}( \delta \tilde{T}  \geq y) \simeq  \overline{ e^{- e^{ \ell \hat{\gamma} - (\ell/v^* + \ell^{\alpha} y) \hat{I}_{{\rm q}}\left( \frac{\ell}{\ell/v^* + \ell^{\alpha} y} \right) +\hat{\lambda}\left( \frac{\ell}{\ell/v^* + \ell^{\alpha} y} \right) ( \ell/v^* + \ell^{\alpha} y)^{\theta_d} \hat \chi}} }\ssp .
\eea
This follows from the idea that once at least a particle has entered $B(\ell)$, some particles will be present there forever (recall that here we are considering a setting with $\ln N/L$ fixed, see also end of the section). 
Defining as before $v^*=v_{\hat \gamma}^*$ the solution of $\hat{\gamma} v^*_{\hat \gamma} = \hat{I}_{{\rm q}}(v^*_{\hat \gamma})$ we get, taking $\alpha = \theta_d$
\bea
{\rm Prob}( \delta \tilde{T}  \geq y) \simeq e^{-e^{ \frac{\ell}{v^*_{\hat \gamma}} \hat{I}_{{\rm q}}'(v^*_{\hat \gamma}) \frac{y(v^*_{\hat \gamma})^2}{\ell^{1-\alpha}}  - \ell^{\alpha}  y \hat{I}_{{\rm q}}(v^*_{\hat \gamma})  + \hat{\lambda}(v^*_{\hat \gamma}) (\ell/v^*_{\hat \gamma})^{\theta_d} \hat{\chi}}}
\eea
This shows that ${\rm Prob}( \delta \tilde{T}  \geq  \frac{y}{\frac{(v^*_{\hat \gamma})^{\theta_d}}{\lambda(v^*_{\hat \gamma})} ( \hat{I}_{{\rm q}}(v^*_{\hat \gamma}) -  v^*_{\hat \gamma} \hat{I}'_{{\rm q}}(v^*_{\hat \gamma}))}) = {\rm Prob}(\hat{\chi} \leq y) $.  Rewriting ${I}_{{\rm q}}(v^*_{\hat \gamma})  -v^*_{\hat \gamma} \hat{I}'_{{\rm q}}(v^*_{\hat \gamma}) =  v^*_{\hat \gamma} (\hat{\gamma}  - \hat{I}'_{{\rm q}}(v^*_{\hat \gamma}) )  =- \frac{(v^*_{\hat \gamma})^2}{v'(\gamma)}$ we obtain ${\rm Prob}( \delta \tilde{T}  \geq - \frac{y}{\frac{(v^*_{\hat \gamma})^{\theta_d+2}}{\lambda(v^*_{\hat \gamma}) (v^*_{\hat \gamma})'(\gamma) }}) = {\rm Prob}(\hat \chi \leq y) $ and we finally recover \eqref{reshitt} as before but with ($(\lambda, \chi) \to (\hat \lambda, \hat \chi)$) to account for the change of geometry. For the finite ball considered above, we thus expect that 
$\hat \chi$ has the universal KPZUC ``point to point" distribution in any $d$. By varying the nature of the domain
one can obtain various subclasses of ``initial conditions" within the KPZUC. 

It is important to comment that in this section and in the previous one we have scaled $N \sim \exp(\gamma t) \sim \exp(\hat \gamma \ell)$. Hence the number density of the cloud of particles is very large, $N/(D t)^{d/2} \gg 1$ within the typical
diffusion volume, but (by definition) becomes of order 1 near the edge of the cloud, on which we focus. Still, by self-consistency, $R$ must be chosen not too small in order for the expected number of particles in $B(\ell)$ at time $T_{{\rm Hit}}(\ell)$ to remain of order $1$. Hence we must have $N e^{-\frac{\ell}{v^*_{\hat \gamma}} I_{{\rm q}}(v^*_{\hat \gamma})} \frac{R^d}{(D \ell/v^*_{\hat \gamma})^{d/2}} \simeq 1$. The choice $N = e^{\hat \gamma \ell}$ takes care of the exponential decay of the probability, and $R$ must be chosen as scaling as $R \sim \sqrt{D \ell/v^*_{\hat \gamma}}$ to take care of the overall dilution factor at the hitting time (read-off here from the averaged probability). Hence $R$ grows with $\ell$ but $R = o(\ell^{\theta_d})$, ensuring that the distribution of $\chi$ remains that of the point-to-point KPZ sub-universality class. Note finally that in this scaling the hitting time distribution does not depend on $R$: the dependence of $T_{{\rm Hit}}(\ell)$ on the size of the ball presumably only appears at higher order in $\ell$.
\end{widetext}
\end{document}